%% file: main.tex
\documentclass[sigconf]{acmart}
\setcopyright{acmcopyright}
\copyrightyear{2021}
\acmYear{2021}
\acmConference[KDD '21] {Proceedings of the 27th ACM SIGKDD Conference on Knowledge Discovery and Data Mining}{August 14--19, 2021}{Virtual Event, Singapore.}
\acmBooktitle{Proceedings of the 27th ACM SIGKDD Conference on Knowledge Discovery and Data Mining (KDD '21), August 14--19, 2021, Virtual Event, Singapore}
\acmPrice{15.00}
\acmISBN{978-1-4503-8332-5/21/08} 
\acmDOI{10.1145/XXXXXX.XXXXXX}
\renewcommand\footnotetextcopyrightpermission[1]{} 

\usepackage{xcolor}
\usepackage[ruled,linesnumbered]{algorithm2e}
\usepackage{graphicx}
\usepackage{subfig}
\usepackage{threeparttable}
\usepackage{booktabs}
\usepackage{multirow}
\usepackage{caption}
\usepackage{url}
\usepackage{diagbox}
\usepackage{svg}
\usepackage{enumitem}
\usepackage{fancyhdr}

\begin{document}
\fancyhead{}

\title{Where are we in embedding spaces? A Comprehensive Analysis on Network Embedding Approaches for Recommender Systems}
\author{Sixiao Zhang}
\authornote{Both authors contributed equally to this research.}
\email{zsx57575@gmail.com}
\affiliation{%
  \institution{University of Technology Sydney}
}

\author{Hongxu Chen}
\authornotemark[1]
\authornote{Corresponding author.}
\email{hongxu.chen@uts.edu.au}
\affiliation{%
  \institution{University of Technology Sydney}
}

\author{Xiao Ming}
\email{201934855@mail.sdu.edu.cn}
\affiliation{%
  \institution{Shandong University}
}

\author{Lizhen Cui}
\email{clz@sdu.edu.cn}
\affiliation{%
 \institution{Shandong University}
}

\author{Hongzhi Yin}
\email{h.yin1@uq.edu.au}
\affiliation{%
  \institution{The University of Queensland}
}

\author{Guandong Xu}
\authornotemark[2]
\email{guandong.xu@uts.edu.au}
\affiliation{%
  \institution{University of Technology Sydney}
}

\renewcommand{\shortauthors}{Zhang, et al.}

\begin{abstract}
Hyperbolic space and hyperbolic embeddings are becoming a popular research field for recommender systems. However, it is not clear under what circumstances the hyperbolic space should be considered. To fill this gap, This paper provides theoretical analysis and empirical results on when and where to use hyperbolic space and hyperbolic embeddings in recommender systems. Specifically, we answer the questions that which type of models and datasets are more suited for hyperbolic space, as well as which latent size to choose. We evaluate our answers by comparing the performance of Euclidean space and hyperbolic space on different latent space models in both general item recommendation domain and social recommendation domain, with 6 widely used datasets and different latent sizes. Additionally, we propose a new metric learning based recommendation method called SCML and its hyperbolic version HSCML. We evaluate our conclusions regarding hyperbolic space on SCML and show the state-of-the-art performance of hyperbolic space by comparing HSCML with other baseline methods.
\end{abstract}
%

\ccsdesc[500]{Information systems~Recommender systems}

\keywords{Recommender Systems; Hyperbolic Space; Node Embeddings}

\thanks{To appear in SIGKDD 2021}
\maketitle
\sloppy
\input{intro}
\input{preliminaries}
\input{hypotheses}
\input{model}
\input{methodology}
\input{experiment}
\input{related-work}

\section{Conclusion}\label{sec:con}
In this paper, we provide a comprehensive analysis on hyperbolic space for recommender systems, comparing the performance of hyperbolic space with Euclidean space in three aspects: method, dataset, and latent size. To the best of our knowledge, this is the first work to address the advantages and disadvantages of hyperbolic space and give comments and suggestions on when and where to use it. Additionally, we propose SCML and its hyperbolic version HSCML, a distance model for social recommendation. Experiments show that hyperbolic space can easily reach a comparable or better performance than existing Euclidean social recommendation methods with a simple distance model HSCML. A customized model for hyperbolic space may outperform all baselines in the future work.
\balance
\bibliographystyle{ACM-Reference-Format}
\bibliography{ref.bib}
\end{document}

%% file: intro.tex
\section{Introduction}
With the rapid development of Internet and world wide web, recommender systems play an important role in modeling user preference and providing personalized recommendation. Because of the underlying graph structure of most real-world user-item interactions, graph-based recommendation has become a popular research field. Latent space models \cite{hoff2002latent} such as matrix factorization based models and metric learning based models are widely used in graph-based recommender systems. They are capable of learning user and item embeddings and utilizing the expressiveness of the latent space to capture the underlying distribution and the latent structure of the data. Most latent space models are built in Euclidean space, the most straightforward latent space consistent with human cognition. However, the ability of embedding methods to model complex patterns is inherently bounded by the dimensionality of Euclidean space \cite{nickel2017poincare}. Euclidean embeddings suffer from a high distortion when embedding scale-free and hierarchical data \cite{chami2019hyperbolic}. The distances of users and items are not likely to be preserved when embedded into Euclidean space. It is necessary to increase the latent size in order to reduce the distortion. However, as the latent size increases, the resources needed to train and store the model also increase.

\begin{figure}[t]
    \subfloat[]{\includegraphics[width=0.25\linewidth,bb=-10 -10 2000 2000]{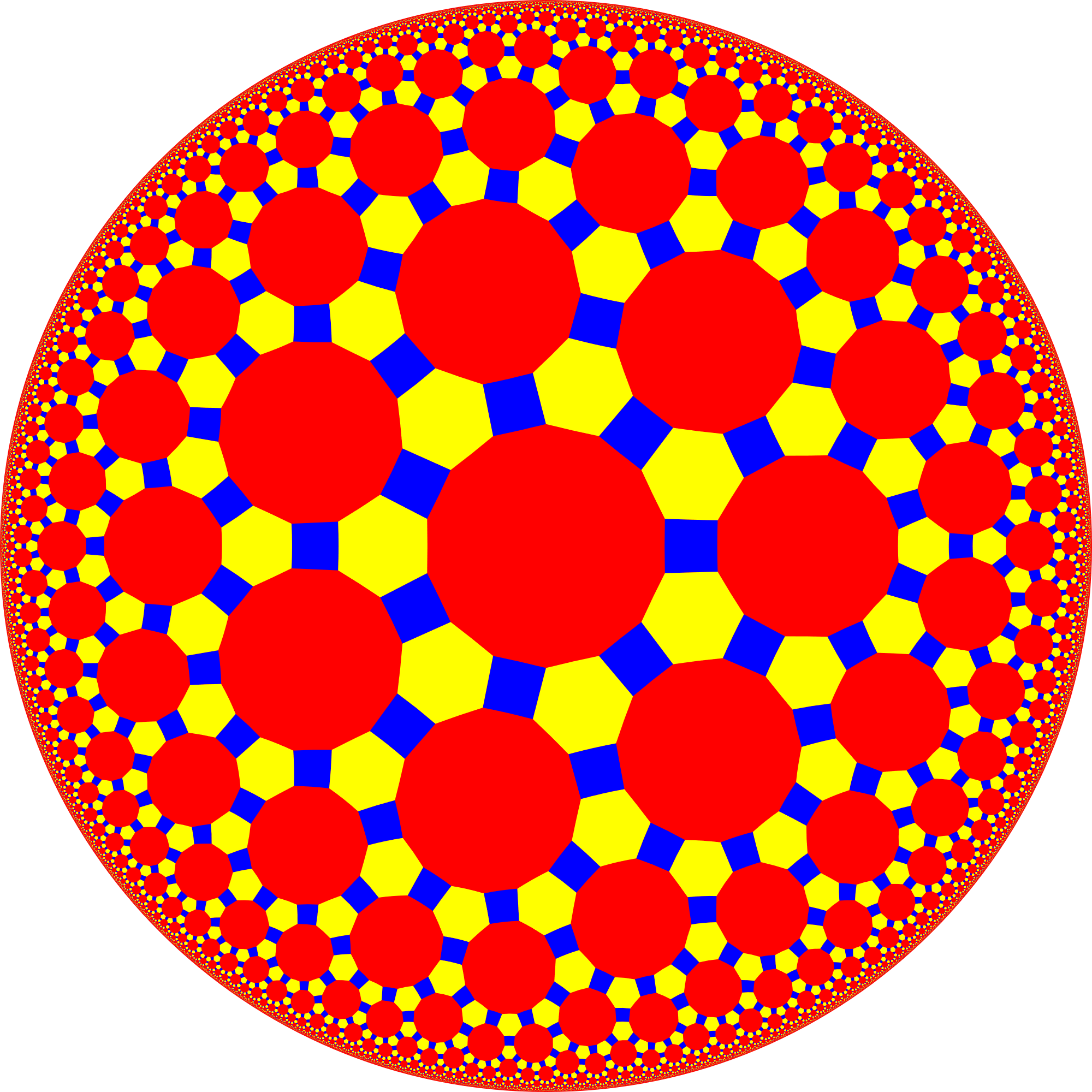}\label{fig:preliminaries a}}
    \hspace{5ex}
    \subfloat[]{\includegraphics[width=0.25\linewidth]{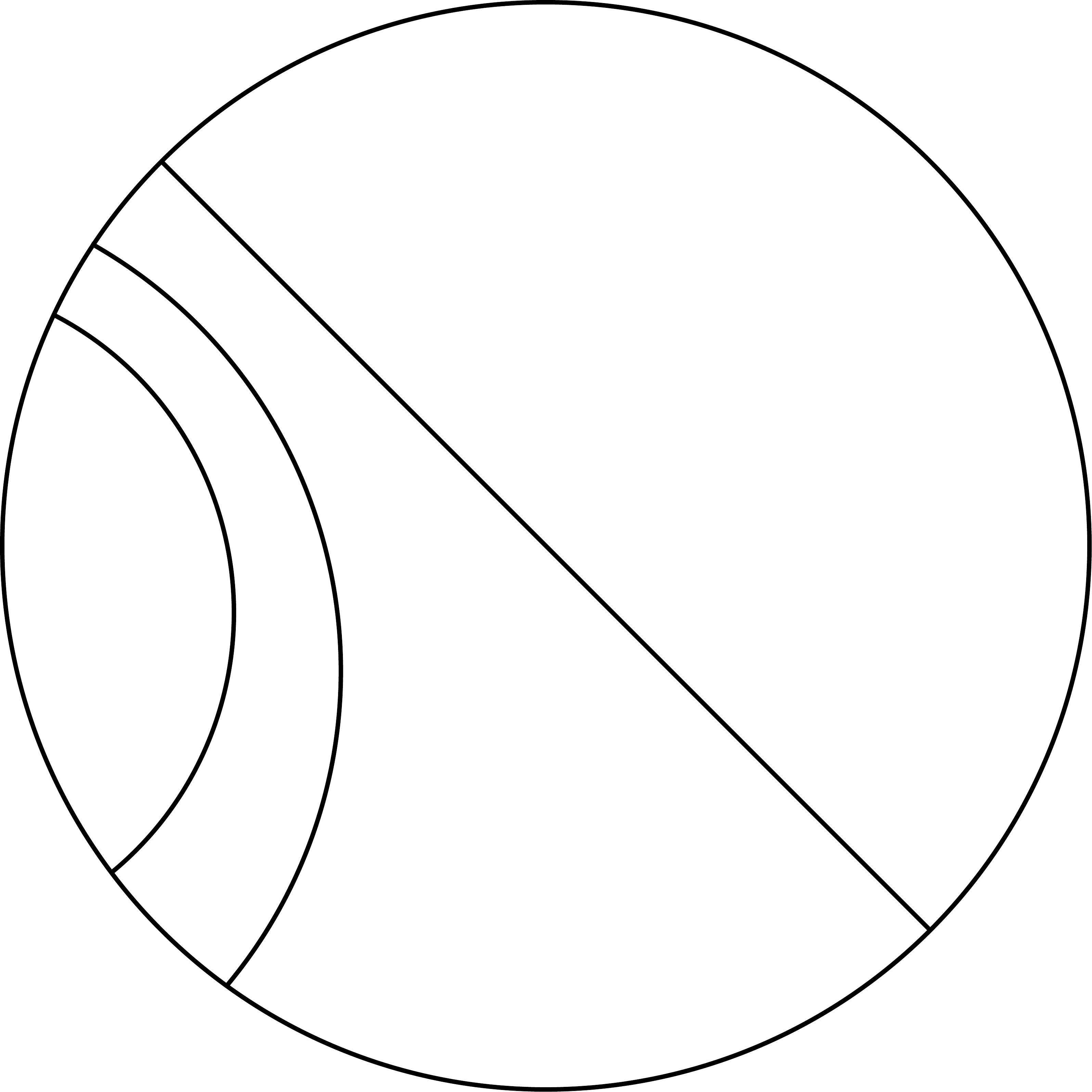}\label{fig:preliminaries b}}
\vspace{-1em}
\caption{(a) Hyperbolic space expands exponentially\protect\footnotemark. (b) 2D Poincaré disk and geodesics.}
\label{fig:preliminaries}
\vspace{-2em}
\end{figure}
\footnotetext{\url{https://commons.wikimedia.org/w/index.php?curid=86312698}}

To solve this problem, a new latent space with a smaller embedding distortion is needed. Recently, hyperbolic space has been explored as a new latent space and has shown impressive performance. It outperforms Euclidean space in many domains including natural language processing \cite{dhingra2018embedding}, link prediction and node classification  \cite{chami2019hyperbolic}, top-n recommendation \cite{vinh2020hyperml}, etc. A key property of hyperbolic space is that it expands faster than Euclidean space. Euclidean space expands polynomially, while hyperbolic space expands exponentially. As shown in \autoref{fig:preliminaries a}, if we treat it as an circle in Euclidean space, the polygons away from the center of the circle are smaller than those close to the center. But if we treat it as a circle in hyperbolic space, then every polygon has the same size. This suggests that, to embed the same group of data, the space needed in Euclidean space is larger than hyperbolic space. In other words, the data capacity of hyperbolic space is larger than Euclidean space. Another important property of hyperbolic space is that it can preserve the hierarchies of data. Many real-world data such as texts, e-commercial networks and social networks exhibit an underlying hierarchical tree-like structure with power-law distribution \cite{adcock2013tree, nickel2017poincare}. Meanwhile, hyperbolic space can be thought as a continuous version of trees. Therefore, such networks are consistent with hyperbolic space due to their analogous structure, and can be naturally modeled by hyperbolic space with a much lower distortion compared to Euclidean space.

Although hyperbolic embeddings are gaining great attention for recommender systems nowadays \cite{liu2019hyperbolic, wang2019hyperbolic, chami2019hyperbolic, chamberlain2019scalable}, it is not clear under what circumstances the hyperbolic space should be considered. A critical analysis and guidance are missing in existing literature. As a current trendy topic of hyperbolic embeddings, practitioners are indiscriminately attempting to transfer variants of existing algorithms into hyperbolic geometry in regardless of having strong motivations. Therefore, it is timely needed for a fair comparison and analysis on existing recommendation algorithm in hyperbolic space. It is unknown whether existing methods perform better in hyperbolic space compared against their original feature spaces. From dataset perspective, if hyperbolic geometry is regarded as a more suitable space for particular recommendation scenarios, we are wondering whether hyperbolic space is versatile across different datasets or is only suitable for particular datasets? Fair latitudinal comparisons are also missing in current existing works. 

Our work aims to fix the gaps and provide a comprehensive analysis on hyperbolic space and hyperbolic embeddings for recommender systems. We first introduce hyperbolic space and the Poincaré ball model. Then we propose three hypotheses regarding the performance of hyperbolic space on different models, datasets, and latent sizes. Meanwhile, we propose a metric learning based social recommendation approach named Social Collaborative Metric Learning (SCML) and its hyperbolic version Hyperbolic Social Collaborative Metric Learning (HSCML). Our implementation is available at Github\footnote{\url{https://github.com/RinneSz/Social-Collaborative-Metric-Learning}}. We empirically validate our hypotheses on 6 benchmark datasets and 6 models from both general item recommendation domain and social recommendation domain. Finally we draw conclusions and give comments on how to use hyperbolic space. Our main contributions are:
\begin{itemize}[leftmargin=1em]
    \item We provide theoretical analysis and empirical results to validate our three hypotheses: hyperbolic space is more suited for distance models than projection models; hyperbolic space is more powerful on datasets with a low density; hyperbolic space greatly outperforms Euclidean space when the latent size is small, but as the latent size increases, Euclidean space becomes comparable with hyperbolic space.
    \item We address the drawbacks of hyperbolic space, and give comments on when and where to use hyperbolic space.
    \item We propose a metric learning based social recommendation method SCML and its hyperbolic version HSCML and validate our hypotheses on them. We also show that hyperbolic space has state-of-the-art performance by comparing HSCML with other baselines on two benchmark datasets.
\end{itemize}

%% file: preliminaries.tex
\section{Preliminaries}
\subsection{The Poincaré Ball Model}
There are five well-known hyperbolic models: the Klein model $\mathbb{K}$, the Poincaré ball model $\mathbb{D}$, the half-plane model $\mathbb{P}$, the hyperboloid model $\mathbb{H}$, and the hemisphere model $\mathbb{J}$. Each of them is defined on a different domain in $\mathbb{R}$, and has a different Riemannian metric. Readers can refer to \cite{cannon1997hyperbolic} for more detailed descriptions for them. 

Among the five models, the Poincaré ball model is a very good choice for learning hyperbolic embeddings because it is well-suited for gradient-based optimization \cite{nickel2017poincare}. The definition domain of the Poincaré ball model with constant curvature $-c=-1$ is
\begin{equation}
\small
\label{eq:poincare domain}
    \mathbb{D}=\{(x_{1},...,x_{n}):x_{1}^{2}+\cdots+x_{n}^{2}<\frac{1}{c}\}
\end{equation}
In $\mathbb{R}^{n}$, it is an n-dimensional open unit ball when $c=1$.

The geodesics in the Poincaré ball, which corresponds to the straight lines in Euclidean space, are circles which are orthogonal to the sphere. \autoref{fig:preliminaries b} gives an illustration of a 2D Poincaré disk and its geodesics. The distance between two points $\textbf{u}, \textbf{v}$ in the Poincaré ball is the length along the geodesic. With constant curvature -1, the distance is calculated by
\begin{equation}
\small
\label{eq:poincare distance}
    d_{\mathbb{D}}(\textbf{u}, \textbf{v})=\text{arcosh}\left(1+\frac{2\|\textbf{u}-\textbf{v}\|^{2}}{(1-\|\textbf{u}\|^{2})(1-\|\textbf{v}\|^{2})}\right)
\end{equation}
Here $\|\cdot\|$ is the Euclidean norm, arcosh is the inverse hyperbolic cosine function. The distance is determined only by the position of $\textbf{u}$ and $\textbf{v}$, and therefore changes smoothly with respect to $\textbf{u}$ and $\textbf{v}$. Besides, there is only one center point in the Poincaré ball, which is exactly the origin, so it is convenient to put the root node at the origin, and the leaves will spread around the root layer by layer, capturing the hierarchical structure of the graph. Furthermore, because all the embeddings are within the unit ball, we don't need to regularize or clip the embeddings as we do in Euclidean space, meanwhile maximizing the use of the whole space. Another advantage of the Poincaré ball model is its conformality, which means that the angles are preserved when transforming Euclidean vectors to the Poincaré ball. This is useful for us to define inner product in the Poincaré ball to evaluate the performance of MF-based methods.

As for the other four hyperbolic models, the hemisphere model is used as a tool for visualising transformations between the other models, and is not often used as a model itself; the distance between two points in the Klein model is related to the two ideal points at the intersections of the unit sphere and the straight line that connects the two target points, so the distance function is not as smooth and easy to calculate as the Poincaré ball; the half-plane model uses the upper half-plane as its definition domain, and all the points that lie on the boundary of the half-plane can be treated as the center, therefore it is necessary to regularize the embeddings as they may spread all over the half-plane; the hyperboloid model also has the issue of regularization since its definition domain includes the infinity, too. Therefore, the Poincaré ball model naturally becomes a good choice for learning hyperbolic embeddings compared to the other hyperbolic models.

\subsection{Latent Space Model}
\label{sec:latent space model}
A key reason why hyperbolic space is suitable for recommendation tasks is that, hyperbolic space and real-world datasets both expand exponentially. This is true because we are looking at the datasets at node level. Therefore, theoretically only models that are capable of learning node embeddings are suitable for hyperbolic space. Such models are called latent space models \cite{hoff2002latent}. Each node is represented by an n-dimensional embedding vector. Recommendations are made by comparing the relations between nodes. Based on the different methods that are used to calculate the relation, latent space models can be split into two categories, namely projection models and distance models. Projection models use the inner product to model the relation. One typical example is matrix factorization. For two node embeddings $\textbf{u}$ and $\textbf{v}$, their relation $r_{\textbf{uv}}$ is
\begin{equation}
\small
\label{eq:mf}
    r_{uv}=\textbf{u}^{\text{T}}\textbf{v}+b
\end{equation}
$b$ is the bias. Usually larger $r_{uv}$ means a closer relationship. The relation calculated by \autoref{eq:mf} is used for downstream pairwise ranking tasks or rating prediction tasks. Distance models use the distance between a pair of node embeddings to model their relation. Metric learning approaches are the most well-known distance models. Typically, the relation between two nodes $\textbf{u}$ and $\textbf{v}$ is
\begin{equation}
\small
    r_{uv}=d_{uv}
\end{equation}
$d_{uv}$ is the distance between $\textbf{u}$ and $\textbf{v}$. Two nodes are more likely to have a link or belong to the same category if they are close.

\subsection{Poincare Ball Optimization and Geometry}
Euclidean gradient descent method can't be directly applied to the hyperbolic space, because the value and direction of the hyperbolic gradient might be different from the Euclidean gradient. One solution is to map the hyperbolic embeddings into Euclidean embeddings, and use Euclidean gradient descent method to optimize the Euclidean embeddings, then map the updated Euclidean embeddings back into the hyperbolic embeddings. The tangent space $T_{x}\mathbb{D}$ of a certain point $x$ in the hyperbolic space $\mathbb{D}$ is often used as the target Euclidean space. The common vector operations such as matrix multiplication and addition are done on the tangent space. Thus, the hyperbolic optimization problem is simplified into a Euclidean optimization problem.

In the following part, we will introduce the necessary mathematical basis for the Poincaré ball model. \textbf{Note that the definition domain and and the distance function are already defined in \autoref{eq:poincare domain} and \autoref{eq:poincare distance}}.

\textbf{Riemannian Metric.} The Riemannian metric of an n-dimensional Poincaré ball at point $x$ is
\begin{equation}
\small
\label{eq:metric}
    g_{x}^{\mathbb{D}}=\lambda_{x}^{2}g^{E},\ \text{where}\ \lambda_{x}=\frac{2}{1-c\|x\|^{2}}
\end{equation}
$g^{E}=\textbf{I}_{n}$ is the Euclidean metric, which is an n-dimensional identity matrix. We recover the Euclidean space when $c=0$. 

\textbf{Hyperbolic Norm.} The norm of a Poincaré embedding \textbf{u} is its distance to the origin. When $c=1$, the norm is
\begin{equation}
\small
    \|\textbf{u}\|_{\mathbb{D}}=\text{arcosh}\left(1+\frac{2\|\textbf{u}\|^{2}}{1-\|\textbf{u}\|^{2}}\right)
\end{equation}
The norm is 0 when $\|\textbf{u}\|=0$, and goes to infinity when $\|\textbf{u}\|$ is close to 1.

\textbf{Hyperbolic Inner Product.} Because the Poincaré ball model is a conformal model, for any two embeddings, their angle in the Poincaré ball is the same as their angle in Euclidean space. Therefore we define the Poincaré inner product as
\begin{equation}
\small
    \langle \textbf{u},\textbf{v}\rangle_{\mathbb{D}}=\|\textbf{u}\|_{\mathbb{D}}\cdot\|\textbf{v}\|_{\mathbb{D}}\cdot\text{cos}\langle \textbf{u},\textbf{v}\rangle,\ \text{where}\ \text{cos}\langle \textbf{u},\textbf{v}\rangle=\frac{\langle \textbf{u},\textbf{v}\rangle}{\|\textbf{u}\|\cdot\|\textbf{v}\|}
\end{equation}
Here $\langle\cdot,\cdot\rangle$ is the Euclidean inner product.

\textbf{Exponential Map \& Logarithmic Map.} The mapping from the tangent space $T_{x}\mathbb{D}$ to the hyperbolic space $\mathbb{D}$ is called the exponential map, and the mapping from the hyperbolic space $\mathbb{D}$ to the tangent space $T_{x}\mathbb{D}$ is call the logarithmic map. Usually the origin $o$ is chosen to be the target point $x$ because of the simplicity and symmetry of the mapping function at $o$. With constant negative curvature $-c$, for $\textbf{t}\in T_{o}\mathbb{D}_{c}$ and $\textbf{u}\in \mathbb{D}_{c}$, the exponential map $\text{exp}_{o}^{c}:T_{o}\mathbb{D}_{c}\rightarrow \mathbb{D}_{c}$ and the logarithmic map $\text{log}_{o}^{c}:\mathbb{D}_{c}\rightarrow T_{o}\mathbb{D}_{c}$ at the origin $o$ are defined as
\begin{equation}
\small
    \text{exp}_{o}^{c}(\textbf{t})=\text{tanh}(\sqrt{c}\|\textbf{t}\|)\frac{\textbf{t}}{\sqrt{c}\|\textbf{t}\|},\ \ \text{log}_{o}^{c}(\textbf{u})=\text{artanh}(\sqrt{c}\|\textbf{u}\|)\frac{\textbf{u}}{\sqrt{c}\|\textbf{u}\|}
\end{equation}

\textbf{Hyperbolic Linear Layer.} In some neural network based methods, we need to change the traditional Euclidean linear layer into hyperbolic linear layer. A typical Euclidean linear layer is composed of three parts: weight multiplication, bias addition, and an activation function at its output. Suppose the weight matrix is $\textbf{W}$, the bias vector is $\textbf{b}$, and the activation function is $\sigma$, the Euclidean linear layer with a Euclidean input vector $\textbf{u}$ can be written as
\begin{equation}
\small
    \textbf{y}=\sigma(\textbf{W}\cdot\textbf{u}+\textbf{b})
\end{equation}
The corresponding hyperbolic linear layer is also composed of the above three parts. Note that, for a hyperbolic linear layer, the input $\textbf{u}$ should also be hyperbolic, but the weight matrix $\textbf{W}$ and bias $\textbf{b}$ should be defined on the tangent space, which means that they are Euclidean. The hyperbolic weight multiplication is defined as
\begin{equation}
\small
\label{eq:hyperbolic weight multiplication}
    \textbf{W}\odot\textbf{u}=\text{exp}_{o}^{c}(\textbf{W}\cdot\text{log}_{o}^{c}(\textbf{u}))
\end{equation}
The hyperbolic bias addition is
\begin{equation}
\small
\label{eq:hyperbolic bias addition}
    \textbf{u}\oplus\textbf{b}=\text{exp}_{o}^{c}(\text{log}_{o}^{c}(\textbf{u})+\textbf{b})
\end{equation}
Using \autoref{eq:hyperbolic weight multiplication} and \autoref{eq:hyperbolic bias addition}, the hyperbolic linear layer is defined as
\begin{equation}
\small
    \textbf{y}=\text{exp}_{o}^{c}(\sigma(\text{log}_{o}^{c}((\textbf{W}\odot\textbf{u})\oplus\textbf{b})))
\end{equation}
where $\sigma$ is the activation function defined in Euclidean space such as ReLU or Sigmoid.



%% file: hypotheses.tex
\section{Hypotheses}
\label{sec:hypotheses}
In this section, we present and explain three hypotheses we make about hyperbolic space. Our experiments are designed to verify the correctness of them.

\textbf{Distance models are more suited for learning hyperbolic embeddings than projection models.} Since the projection models optimize the inner product and the distance models optimize the distance, the distribution of the embeddings is different after convergence. For the projection models, the relation between nodes is mainly determined by the angle. Angles of positive pairs are small and angles of negative pairs are obtuse. However, for the distance models, the relation is only determined by the distance. The positive pairs are pushed close to each other, and the negative pairs are pushed away from each other.

If we apply hyperbolic space to projection models, one issue is that, projection models usually regularize the norm of the embeddings in a limited scale, which makes it impossible to push the embeddings far from the center. So in hyperbolic space, the norm of the embedding will also be in a limited scale, making it hard to make use of the outer part of the hyperbolic space where the capacity is the largest efficiently, thus reducing the expressiveness. However, distance models do not have such limitation and can make use of the entire space. Embeddings can be pushed arbitrarily far from the origin as long as the precision allows. Another advantage of distance models is the ability to learn hierarchical information. Nodes on a circle have the same relation with the center node, so a group of leaf nodes will be likely to spread around a root node layer by layer in the form of concentric circles. As we have mentioned that the number of leaf nodes and the capacity of the hyperbolic space are all increasing exponentially, it's easier for distance models to learn low-distortion hyperbolic embeddings compared to projection models. Therefore, distance models should be a better choice for learning hyperbolic embeddings rather than projection models.

\textbf{Hyperbolic space is more powerful compared to Euclidean space when the density of the dataset is small.} Hyperbolic space is naturally capable of extracting the hierarchical information from datasets. It can use such hierarchy to help learn the node embeddings more effectively. Therefore, it can have a relatively good performance even the density is low. However, it is difficult for Euclidean space to extract the hierarchical information. Instead, Euclidean space needs more user-item pairs to help reach a comparable performance with hyperbolic space. That is to say, if the density of the dataset is small, then hyperbolic space is likely to outperform Euclidean space, but if the density is large, Euclidean space can have a comparable performance with hyperbolic space.

\textbf{The performance of hyperbolic space is better than Euclidean space when the latent size is small, but as the latent size increases, the performance of Euclidean space becomes comparable with hyperbolic space.} The expressiveness of hyperbolic space is much larger than Euclidean space. For example, when embedding one type of hierarchy, a 2D Poincaré disk can have arbitrary small distortion, but a 2D Euclidean space will always have some distortion \cite{chen2013hyperbolicity}. So to fully capture all the hierarchical information in the dataset, the number of dimensions needed in hyperbolic space is much smaller than Euclidean space. This is our reason why hyperbolic space should outperform Euclidean space when the latent size is small. On the other side, as the latent size increases, the performance of Euclidean space and hyperbolic space both have an upper bound. Besides, they should share the same upper bound, because both of them are capable of embedding the nodes with an arbitrary low distortion with a large enough latent size. So as the latent size increases, the gap between their performances should decrease. Eventually Euclidean space will have a similar performance with hyperbolic space.

%% file: model.tex
\vspace{-1em}
\section{Evaluating Hyperbolic Space}
\label{sec:evaluatin hyperbolic space}

We evaluate the performance of hyperbolic space in two different recommendation domains, namely general item recommendation and social recommendation. The reasons why these two domains are chosen are that 1) data sparsity issues and long-tailed distributions are common in these two recommendation tasks, which are regraded as more suitable for hyperbolic embedding space. To be specific, the user-item rating distribution is usually power-law in both domains. Moreover, in social recommendation, the user-user social network distribution is also usually power-law. So hyperbolic space could be a more consistent latent space for both general item recommendation and social recommendation. A thorough evaluation on these two tasks will help following-up research avoid potential trial and errors in developing new research directions. 2) general recommendation and social recommendation are most popular tasks in recommender system research area, studying these two classical tasks will have broader impact and value to the community. Most existing baseline approaches have only reported their performance in Euclidean space, therefore we implement them in hyperbolic space and report their performance in both Euclidean space and hyperbolic space on several benchmark datasets. We aim to verify our hypotheses proposed in \autoref{sec:hypotheses} empirically.

\subsection{\textbf{Baselines}}
\label{sec:baselines}

\begin{table}[h]
\vspace{-1.5em}
\small
\begin{center}
 \begin{tabular}{|c|c|c|c|} 
 \hline
 Baseline & Domain & Task & Model\\
 \hline
 MF-BPR & general & top-n & projection \\
 \hline
 CML & general & top-n & distance \\
 \hline
 DMF & general & top-n & projection \\ 
 \hline
 TrustSVD & social & rating & projection \\ 
 \hline
 SoRec & social & rating & projection \\
 \hline
\end{tabular}
\caption{Baseline categories.}
\label{table:baseline categories}
\end{center}
\vspace{-4em}
\end{table}

We conduct experiments on some of the most representative distance models and projection models in general item recommendation domain and social recommendation domain.
\begin{itemize}[leftmargin=1em]
\item \textbf{Matrix Factorization with Bayesian Personalized Ranking (MF-BPR)} \cite{rendle2012bpr}: a matrix factorization based method optimized by Bayesian personalized ranking loss.
\item \textbf{Collaborative Metric Learning (CML)} \cite{hsieh2017collaborative}: a metric learning approach that learns a joint metric space to encode not only users' preferences but also the user-user and item-item similarity.
\item \textbf{Deep Matrix Factorization (DMF)} \cite{xue2017deep}: a novel matrix factorization model using neural network architecture to learn the node embeddings.
\item \textbf{TrustSVD} \cite{guo2015trustsvd}: a social recommendation method that extends SVD++ by incorporating social trust information to help predict user preference.
\item \textbf{SoRec \cite{ma2008sorec}}: a factor analysis approach based on probabilistic matrix factorization which exploits both the user-item rating matrix and the adjacency matrix of the social networks. 
\end{itemize}

\begin{table*}[t]
\small
\begin{center}
 \begin{tabular}{|c|c|c|c|c|c|c|} 
 \hline
 Dataset & \# Users & \# Items & \#  Ratings & Rating Density & \# Social Connections & Social Density\\
 \hline
 Movielens 1M & 6040 & 3706 & 1000209 & 4.4684\% & \diagbox[width=10em,height=1em]{}{} & \diagbox[width=10em,height=1em]{}{}\\
 \hline
 Movielens 100K & 943 & 1682 & 100000 & 6.3047\% & \diagbox[width=10em,height=1em]{}{} & \diagbox[width=10em,height=1em]{}{}\\
 \hline
 Automotive & 2928 & 1835 & 20473 & 0.3810\% & \diagbox[width=10em,height=1em]{}{} & \diagbox[width=10em,height=1em]{}{} \\ 
 \hline
 Cellphones & 27879 & 10429 & 194439 & 0.0669\% & \diagbox[width=10em,height=1em]{}{} & \diagbox[width=10em,height=1em]{}{} \\ 
 \hline
 Epinions & 40163 & 139738 & 664824 & 0.0118\% & 487183 & 0.0302\%\\
 \hline
 Ciao & 10420 & 111520 & 296558 & 0.0255\% & 128797 & 0.1186\%\\
 \hline
\end{tabular}
\caption{Dataset statistics.}
\label{table:datasets statistics}
\end{center}
\vspace{-4em}
\end{table*}

The domains and tasks of these baselines and whether they are distance models or projection models are described in \autoref{table:baseline categories}. We use these five models because of two reasons. First, they are some of the most standard and basic models of their categories. We can compare the performance of Euclidean space and hyperbolic space with little noise or bias which might be introduced by auxiliary model components. MF-BPR is a standard matrix factorization based projection model. CML is a baseline metric learning based distance model. DMF is one of the simplest projection models that incorporate neural network structure. Another reason why we use these five models is that, the number of existing latent space models is limited. Many other popular recommendation methods such as Neural Collaborative Filtering (NeuMF) \cite{he2017neural}, Factorization Machines (FM) \cite{rendle2010factorization}, and GraphRec \cite{fan2019graph} do not belong to latent space models. We have explained in \autoref{sec:latent space model} that, theoretically, only latent space models are suitable for hyperbolic space. For those which are not latent space models, hyperbolic space is not likely to outperform Euclidean space. Even if they happen to outperform Euclidean space, it is difficult to explain the underlying reason because they do not present any hierarchical property.

When we are investigating existing social recommendation methods, we notice that most methods are projection models and are designed for rating prediction tasks. There is a lack in top-n social recommendation with distance models. Therefore, in the next section, we will propose and evaluate a new method to fill this gap.

\subsection{\textbf{Datasets}}
We use four datasets for general item recommendation and two for social recommendation.
\begin{itemize}[leftmargin=1em]
	\item \textbf{MovieLens\footnote{\url{https://grouplens.org/datasets/movielens/}}}: a widely adopted benchmark dataset in the application domain of recommending movies to users provided by GroupLens research. We use two configurations, namely \textbf{MovieLens 1M} and \textbf{MovieLens 100K}.
	\item \textbf{Amazon review data} \cite{he2016ups}: a popular benchmark dataset that contains product reviews and metadata from Amazon. We use two subsets of it, namely \textbf{Cellphones} and \textbf{Automotive}. Both are 5-core subsets\footnote{\url{http://jmcauley.ucsd.edu/data/amazon/}}.
	\item \textbf{Epinions\footnote{\url{http://www.trustlet.org/downloaded_epinions.html}}}: a popular dataset from a consumer review website called Epinions\footnote{\url{http://www.epinions.com}}, where users can rate items and add other users to their trust lists.
	\item \textbf{Ciao\footnote{\url{https://www.cse.msu.edu/~tangjili/datasetcode/}}}: another popular social recommendation dataset from social networking website Ciao\footnote{\url{http://www.ciao.co.uk}}. It also allows users to rate items and trust other users.
\end{itemize}
\autoref{table:datasets statistics} show the statistics of the datasets. Movielens 1M and Movielens 100K have a very large density, about 4\% and 6\%, while other four datasets are all below 0.4\%. Conclusions about the influence of density can be made by comparing the performance of them.

\begin{figure*}[t]
\centering
	\subfloat[Movielens 1M HR]{\includegraphics[width=0.25\linewidth]{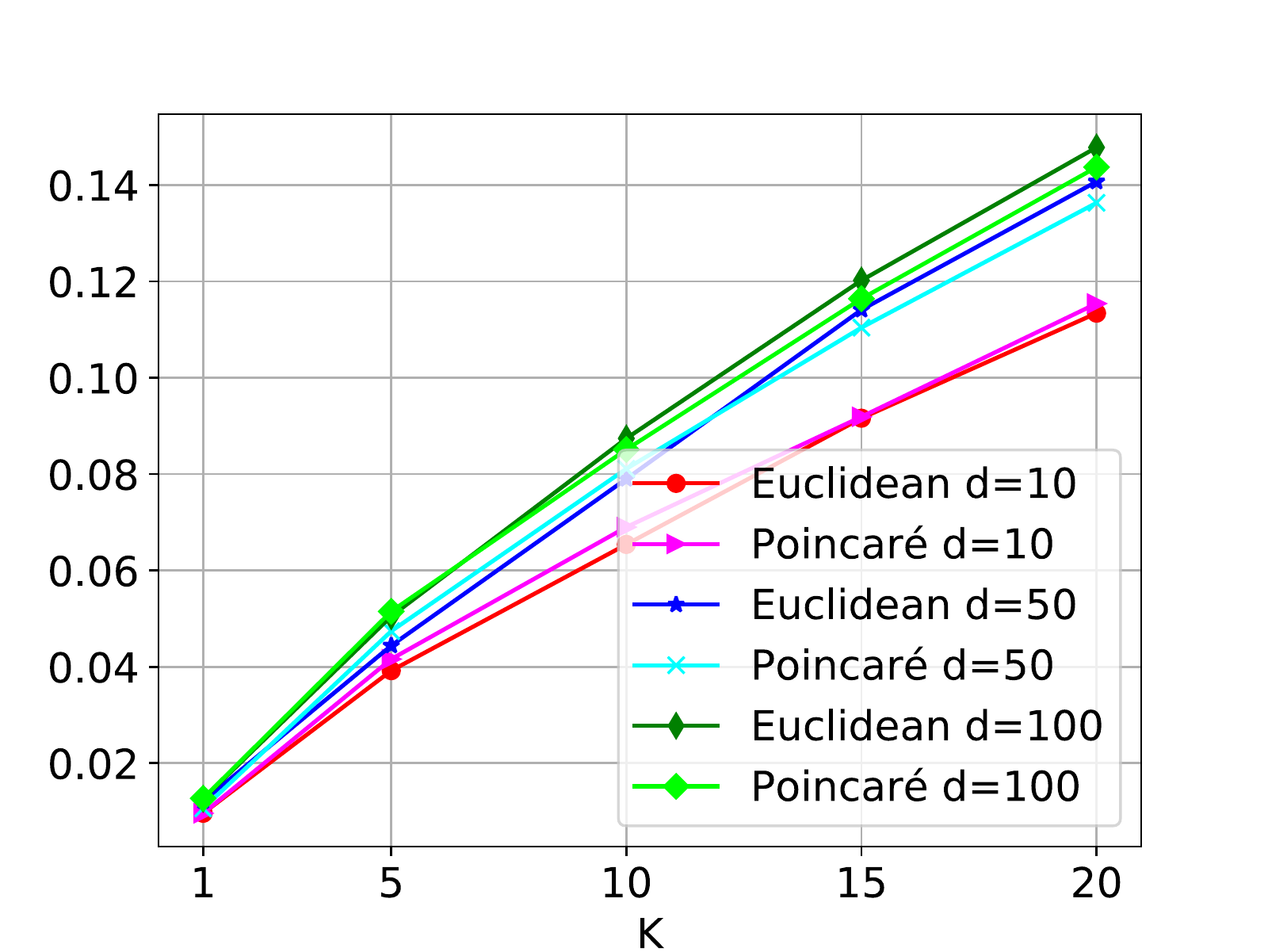}\label{fig:CML a}}
	\subfloat[Movielens 1M NDCG]{\includegraphics[width=0.25\linewidth]{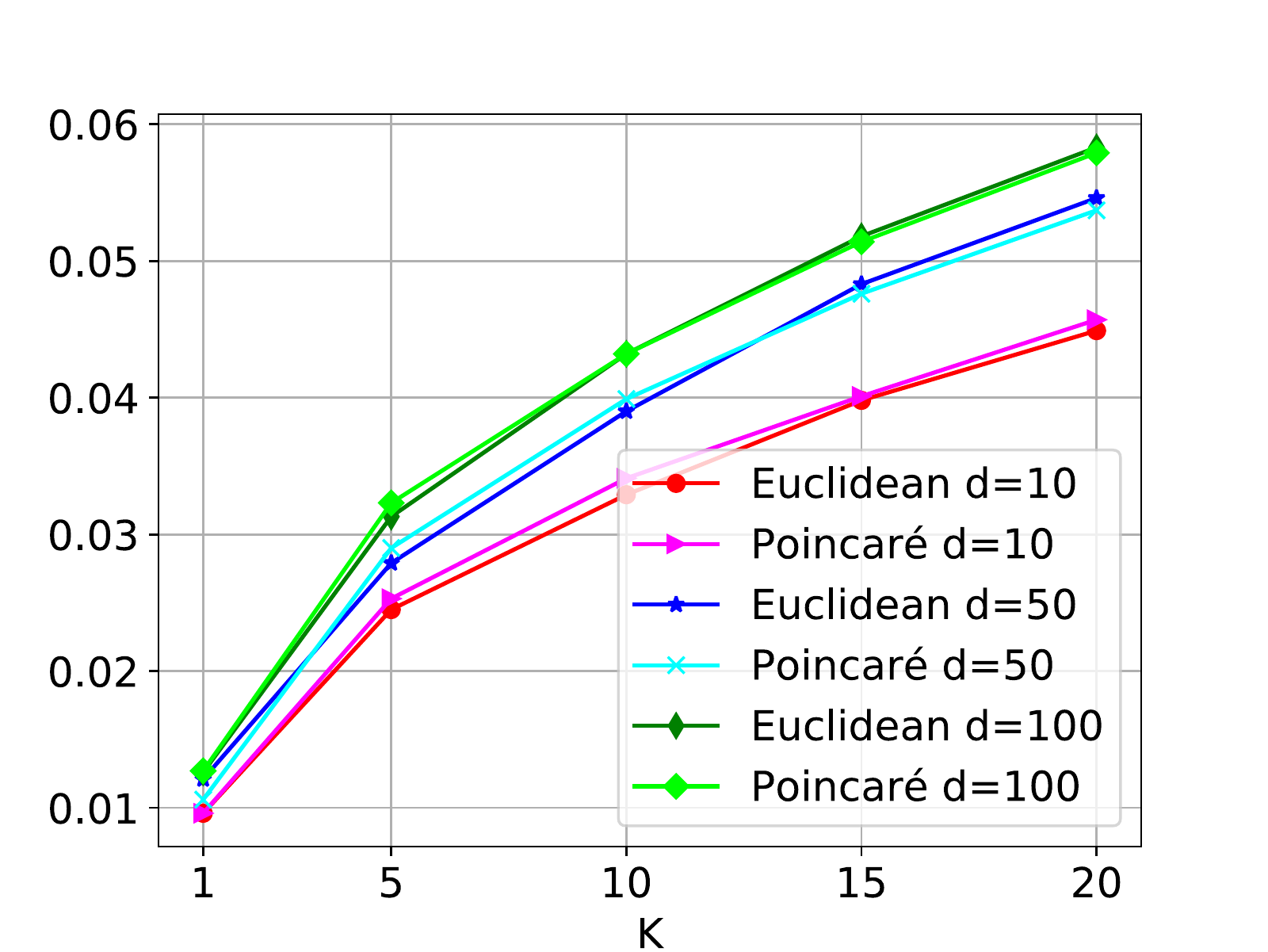}\label{fig:CML b}}
	\subfloat[Movielens 100K HR]{\includegraphics[width=0.25\linewidth]{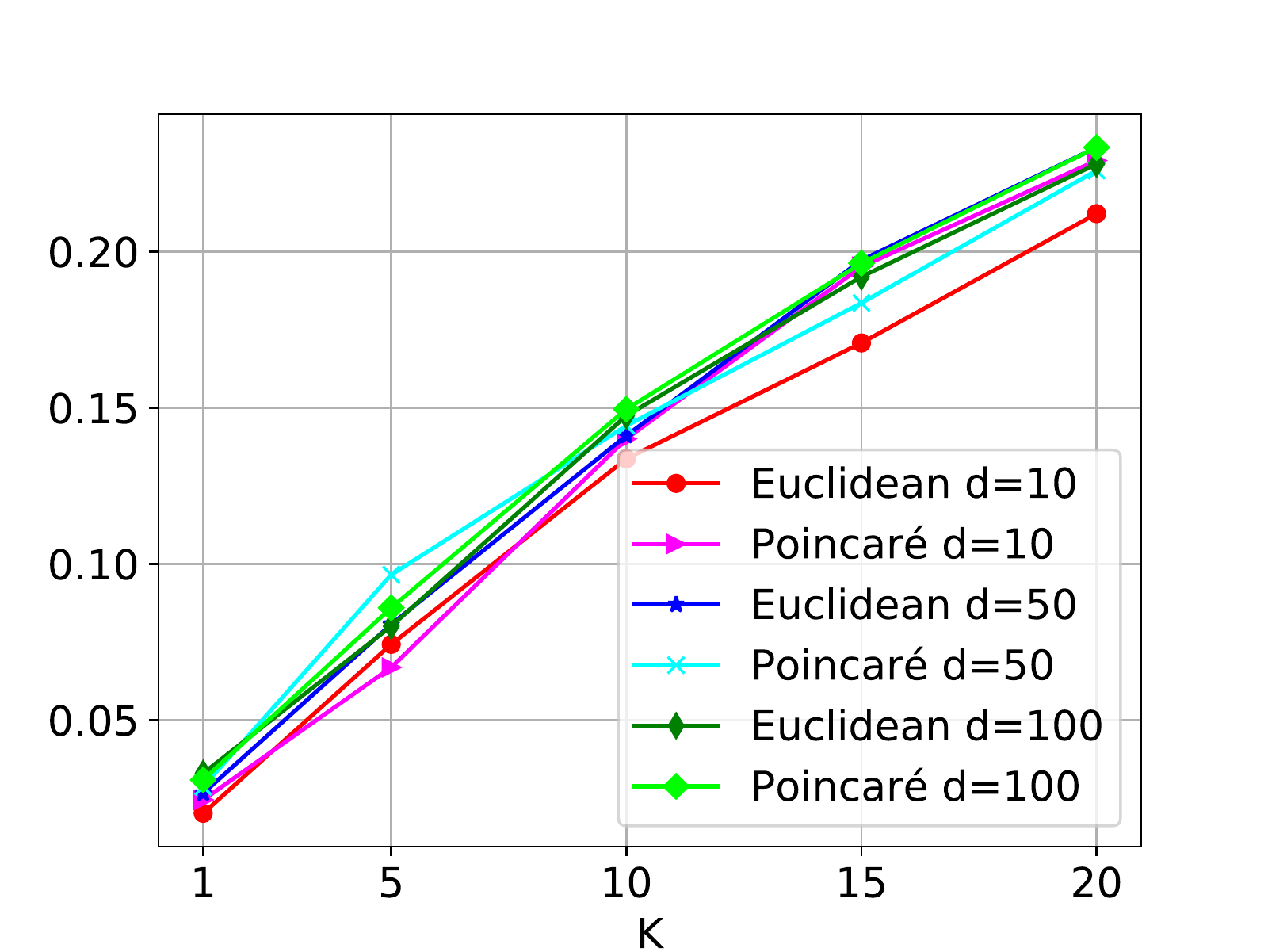}\label{fig:CML c}}
	\subfloat[Movielens 100K NDCG]{\includegraphics[width=0.25\linewidth]{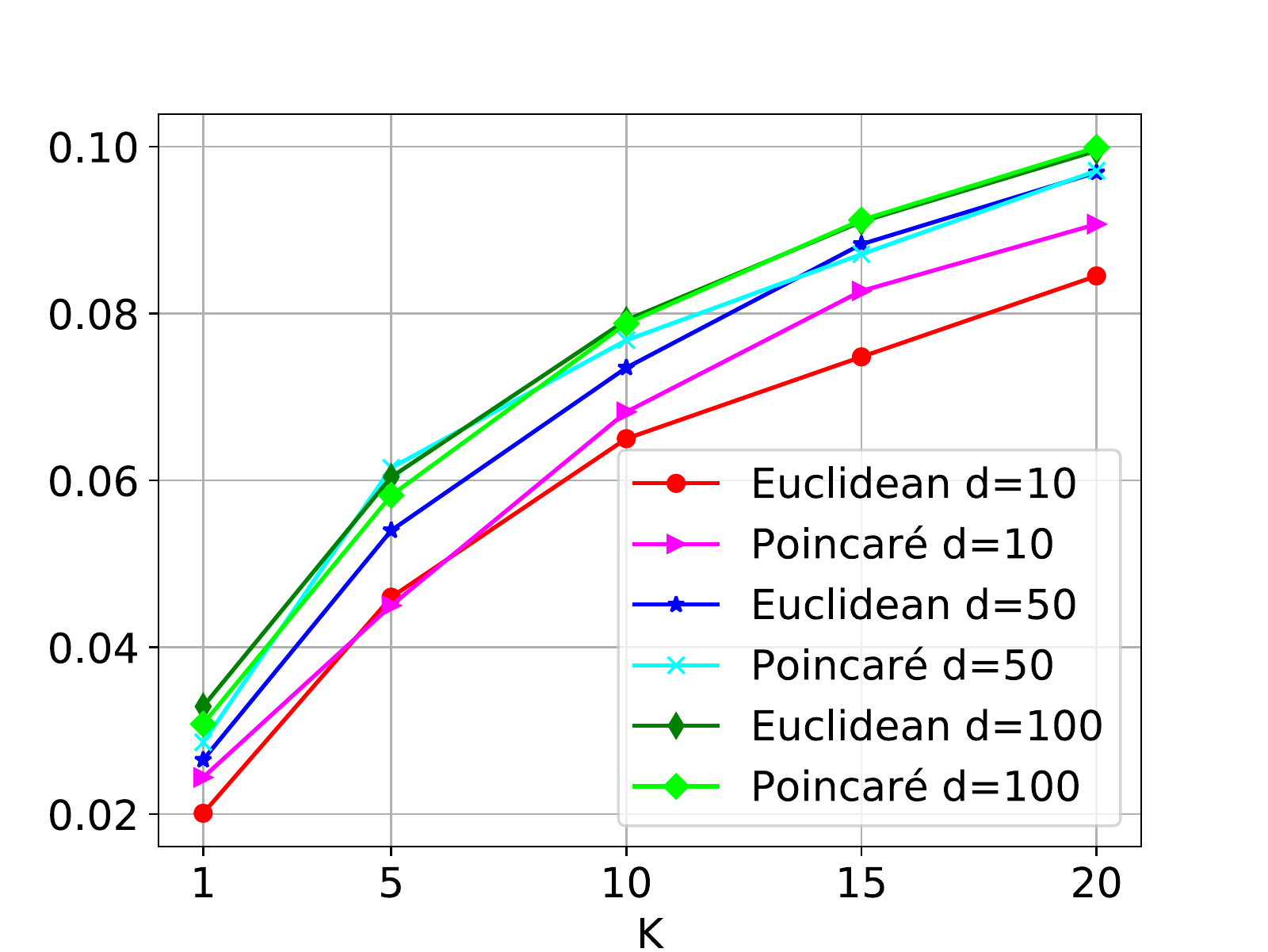}\label{fig:CML d}}\\
	\vspace{-1.3em}
	\subfloat[Automotive HR]{\includegraphics[width=0.25\linewidth]{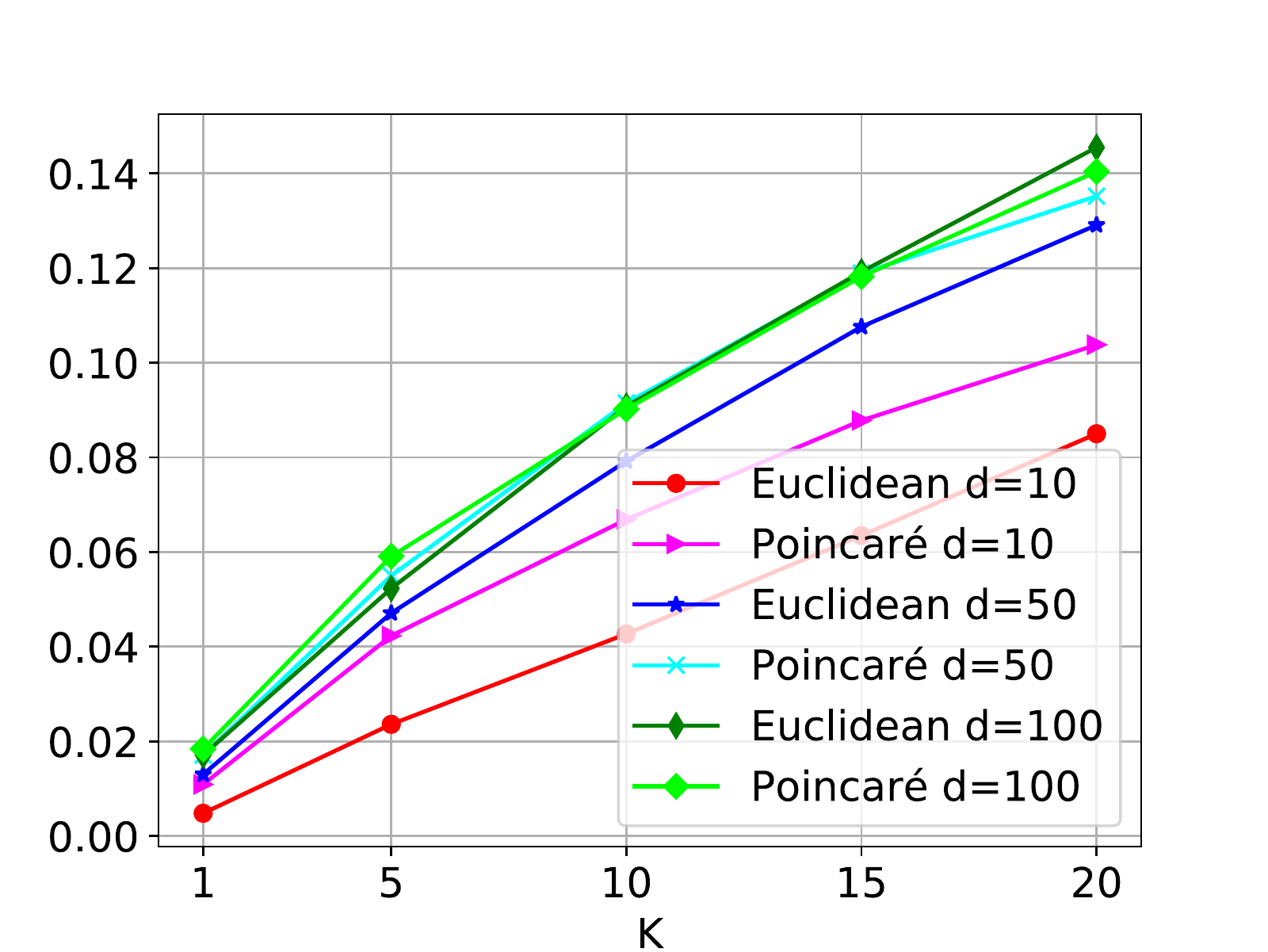}\label{fig:CML e}}
	\subfloat[Automotive NDCG]{\includegraphics[width=0.25\linewidth]{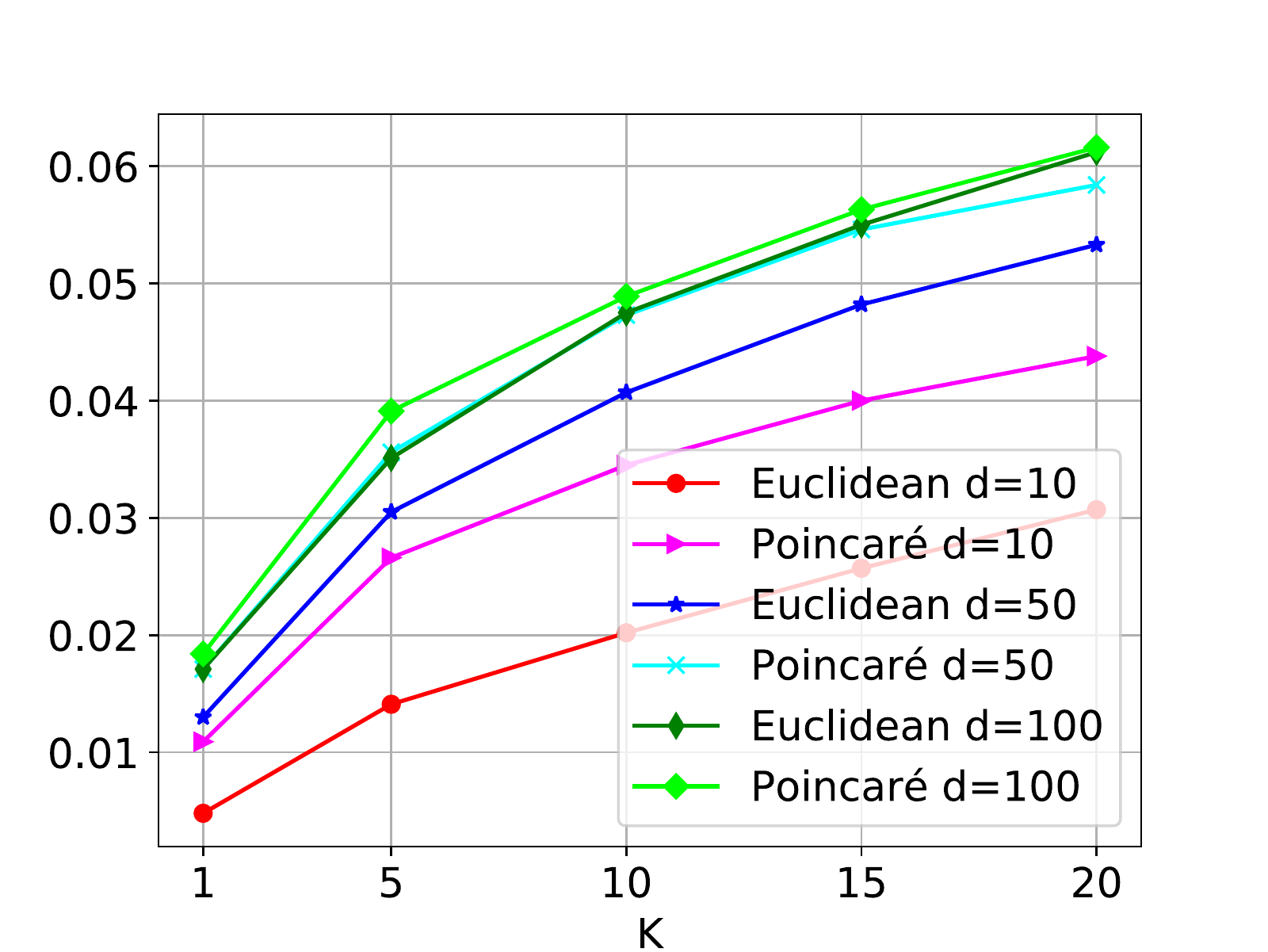}\label{fig:CML f}}
	\subfloat[Cellphones HR]{\includegraphics[width=0.25\linewidth]{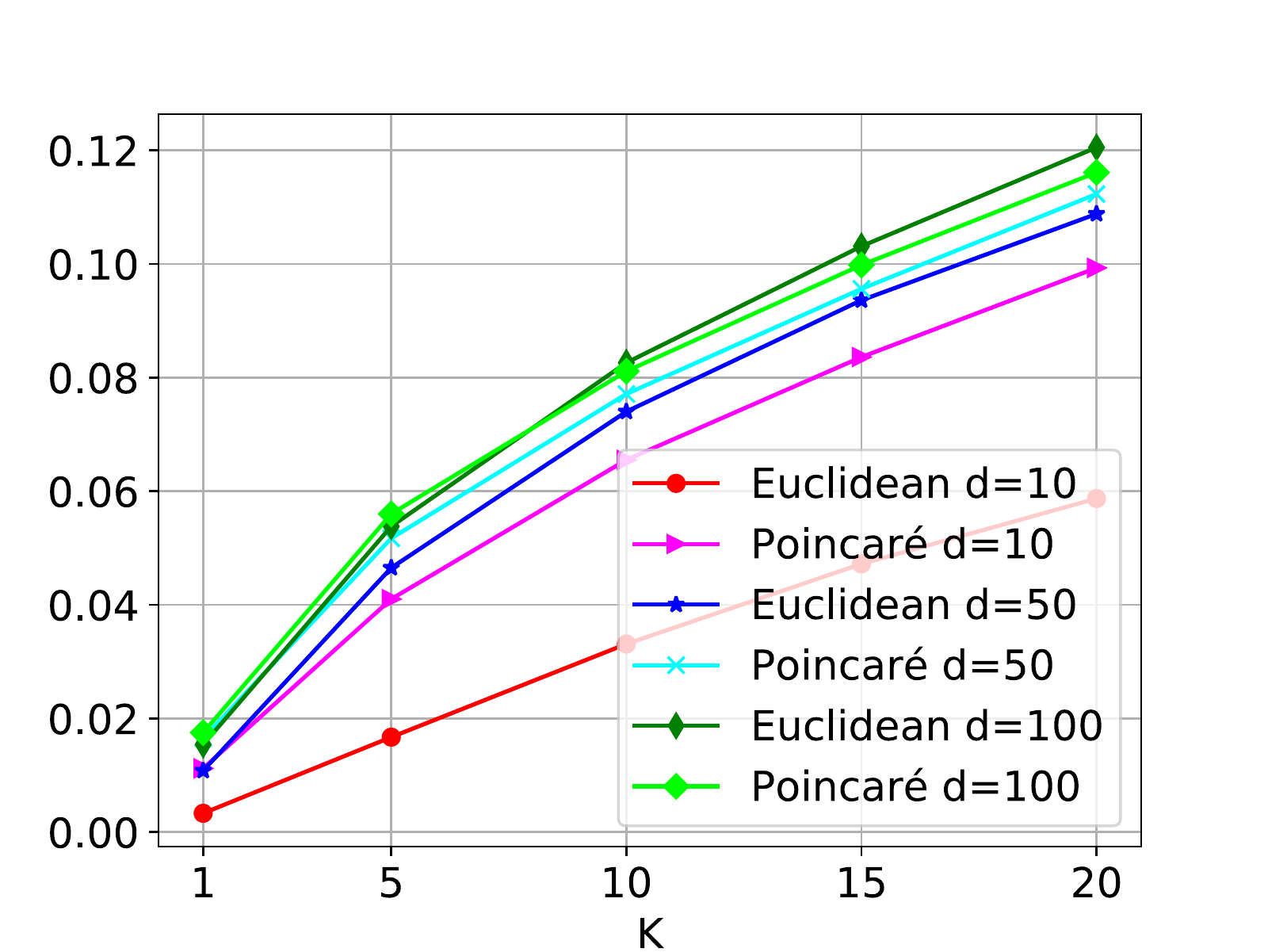}\label{fig:CML g}}
	\subfloat[Cellphones NDCG]{\includegraphics[width=0.25\linewidth]{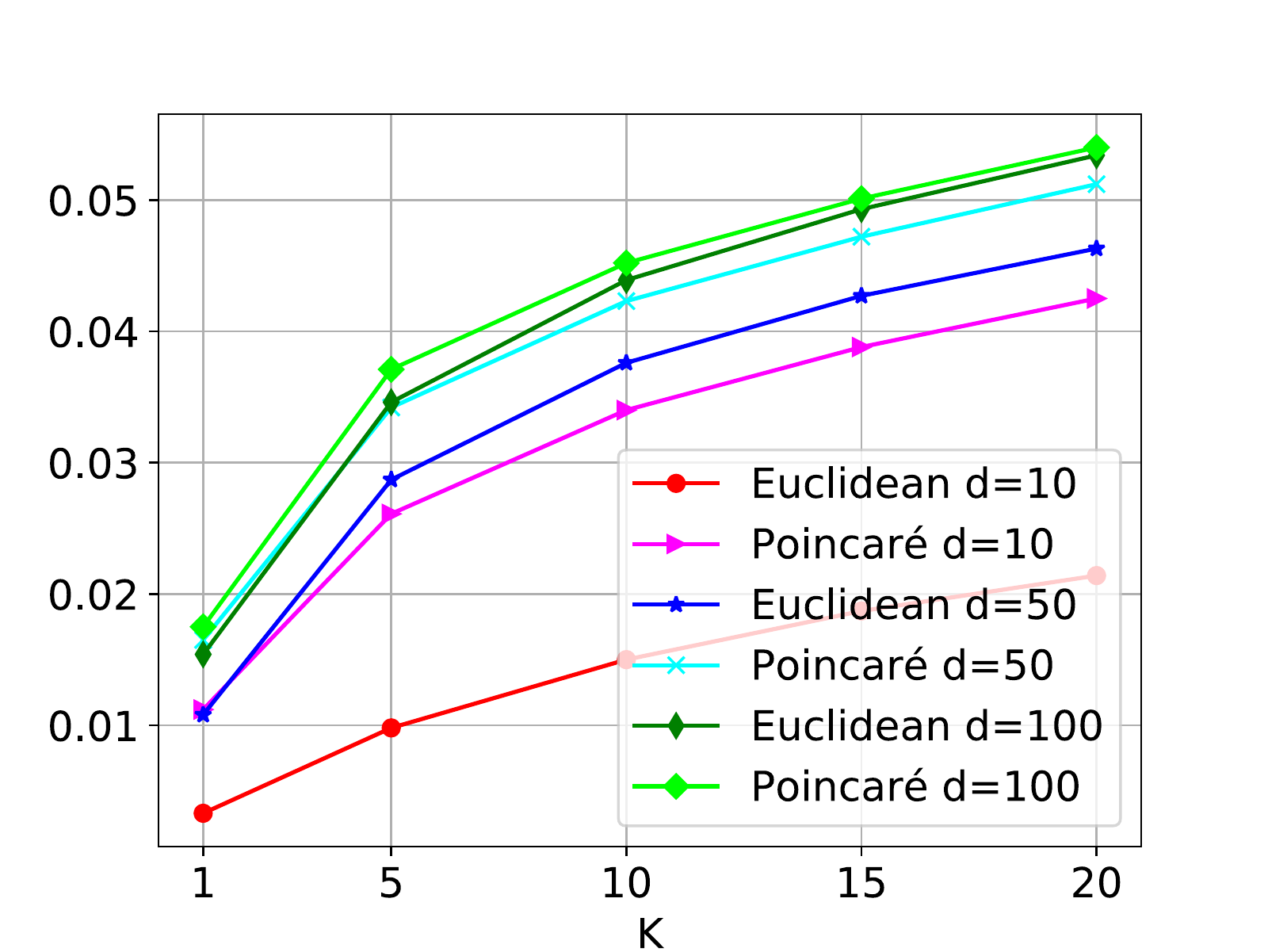}\label{fig:CML h}}
\vspace{-1.3em}
\caption{CML HRs and NDCGs on four datasets.}
\label{fig:CML}
\vspace{-2.5em}
\end{figure*}

\begin{figure*}[t]
\centering
    \subfloat[Movielens 1M HR]{\includegraphics[width=0.25\linewidth]{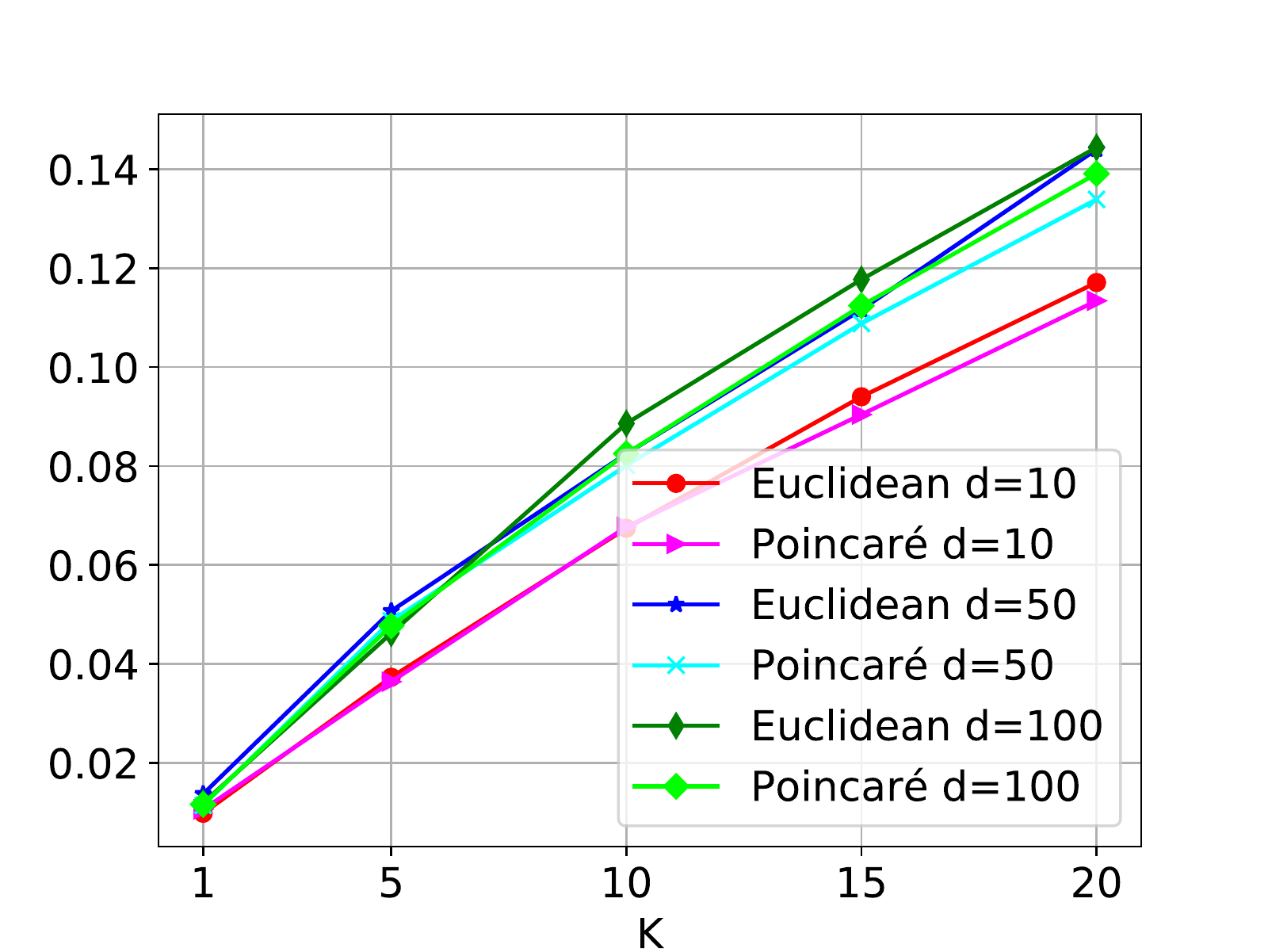}\label{fig:BPR a}}
	\subfloat[Movielens 1M NDCG]{\includegraphics[width=0.25\linewidth]{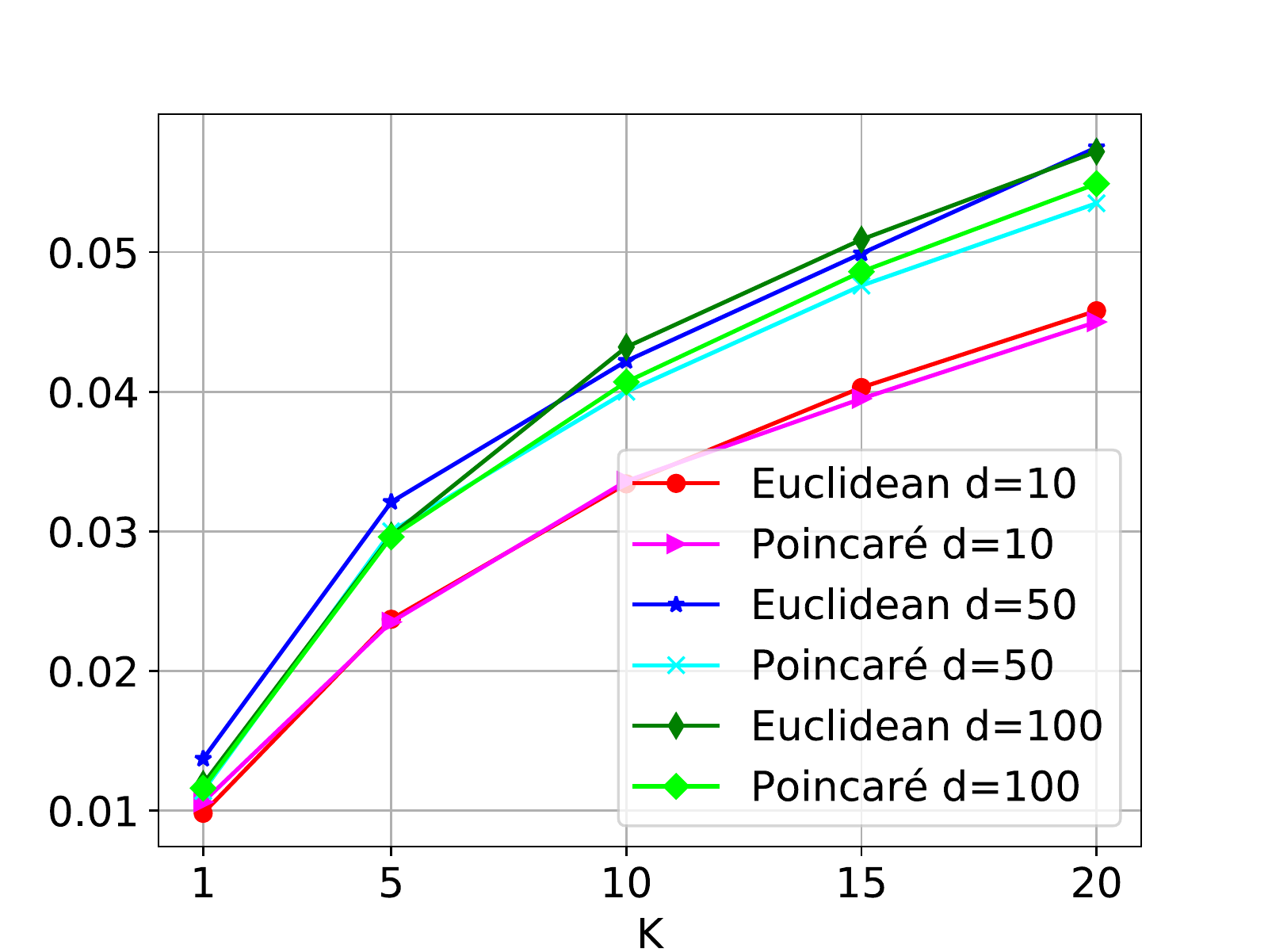}\label{fig:BPR b}}
	\subfloat[Movielens 100K HR]{\includegraphics[width=0.25\linewidth]{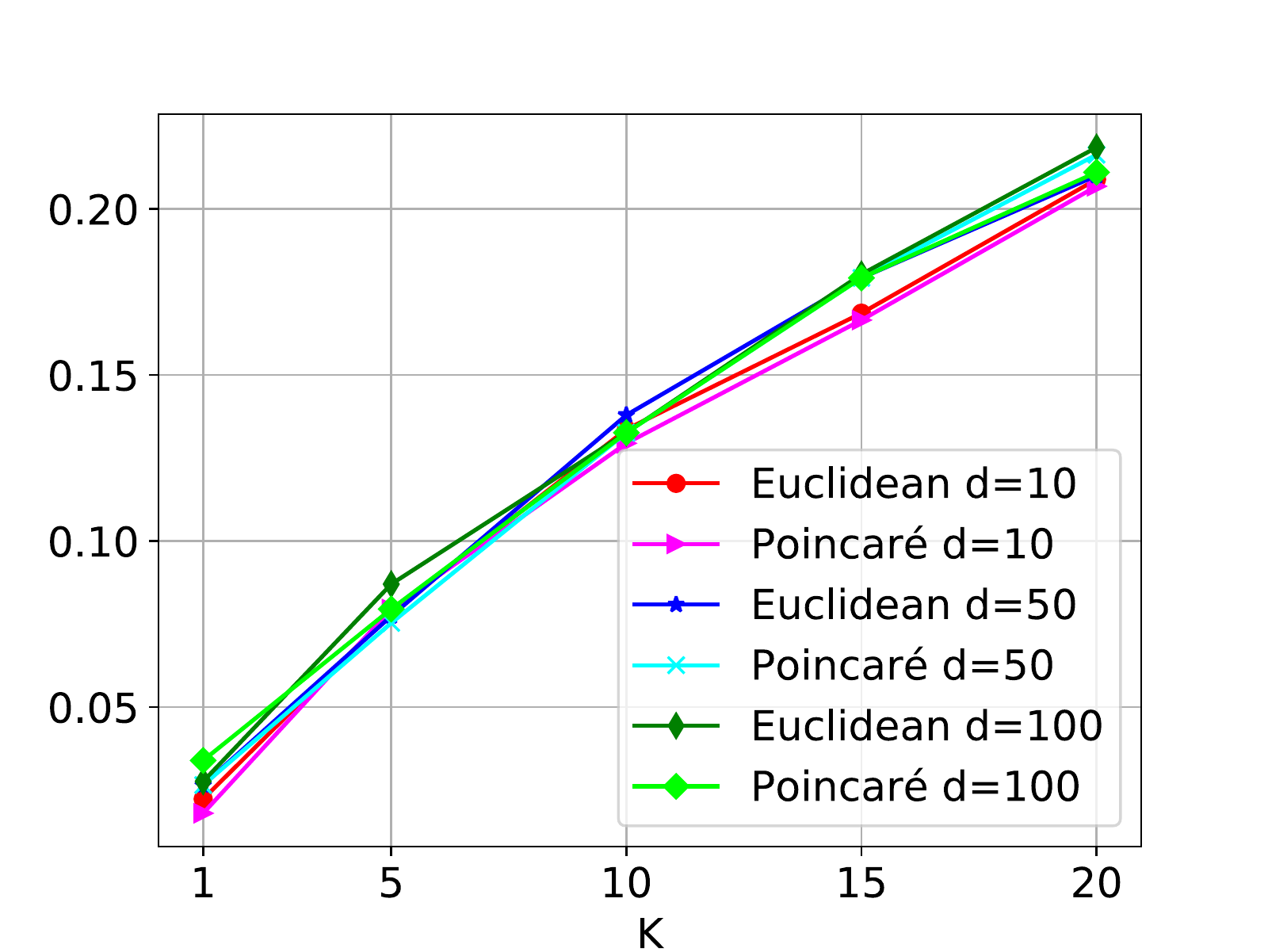}\label{fig:BPR c}}
	\subfloat[Movielens 100K NDCG]{\includegraphics[width=0.25\linewidth]{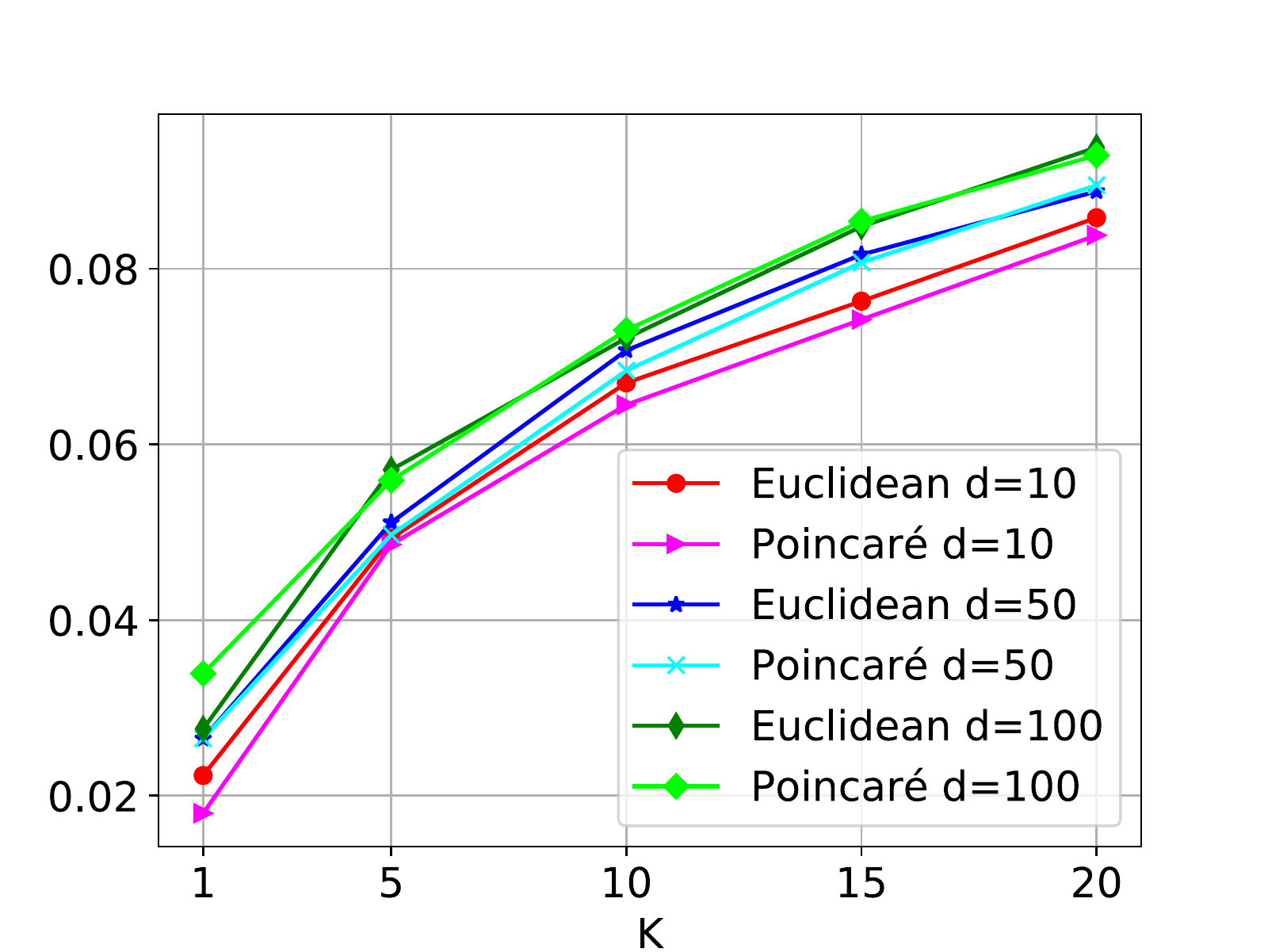}\label{fig:BPR d}}\\
	\vspace{-1.3em}
	\subfloat[Automotive HR]{\includegraphics[width=0.25\linewidth]{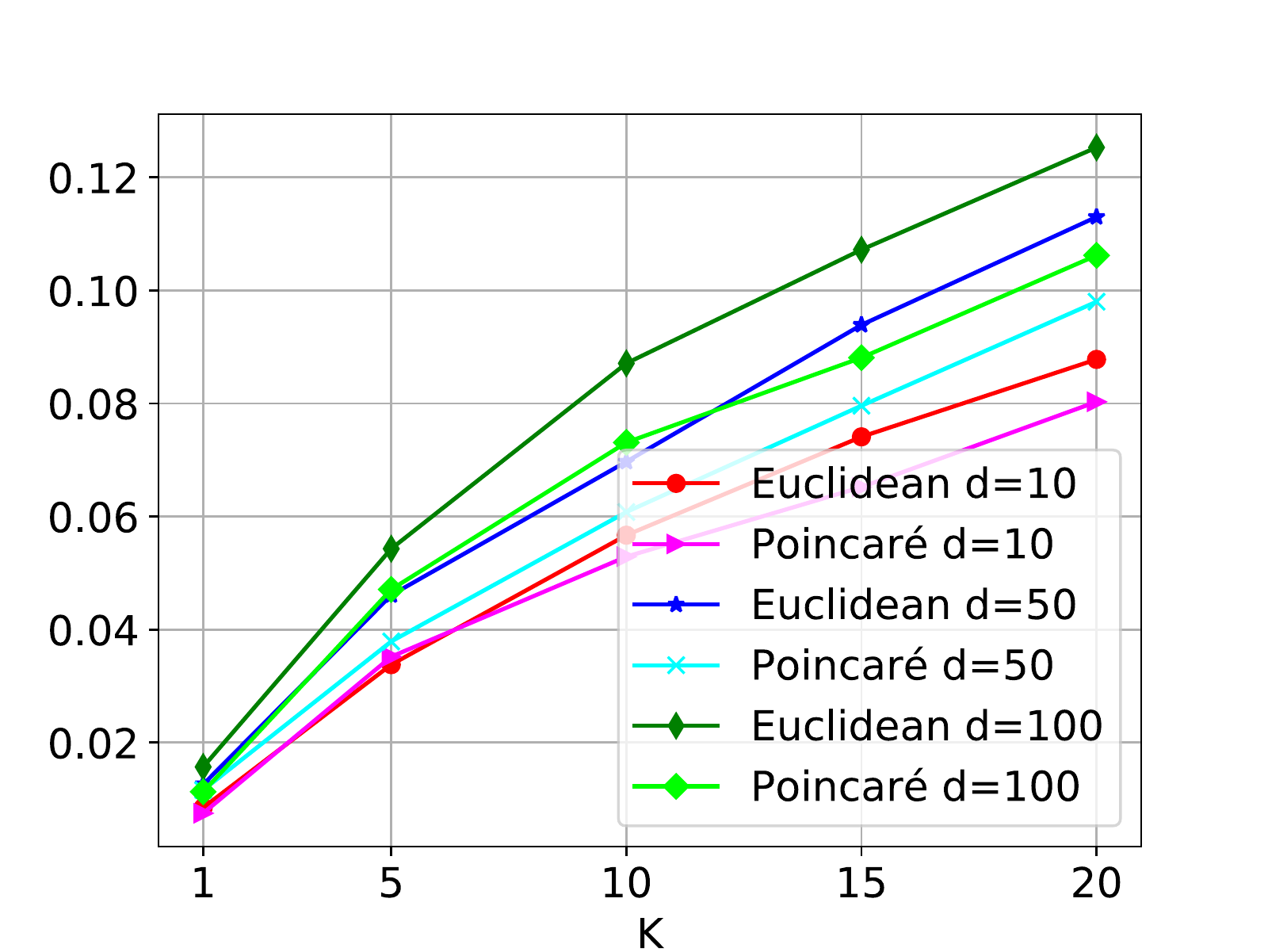}\label{fig:BPR e}}
	\subfloat[Automotive NDCG]{\includegraphics[width=0.25\linewidth]{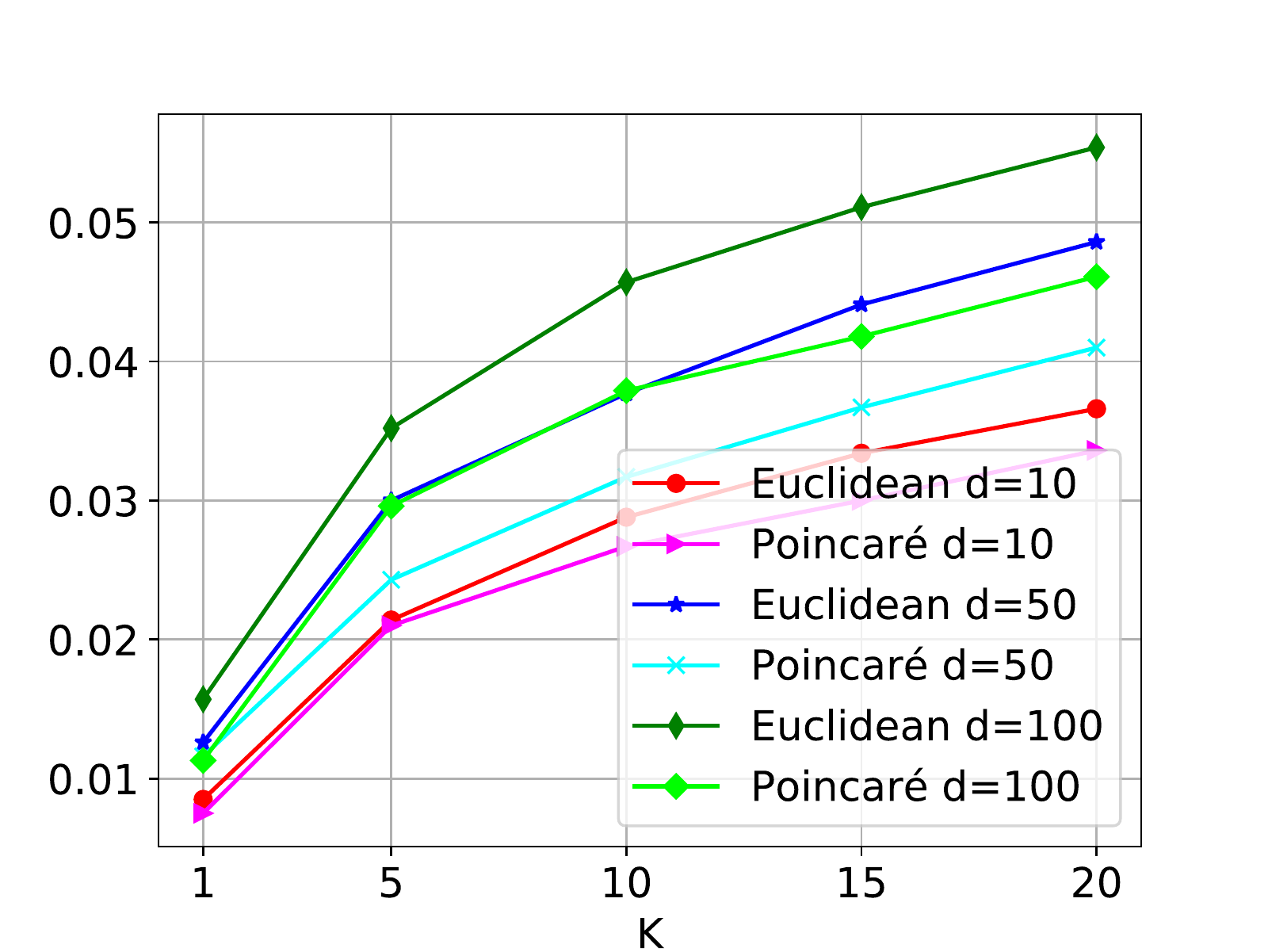}\label{fig:BPR f}}
	\subfloat[Cellphones HR]{\includegraphics[width=0.25\linewidth]{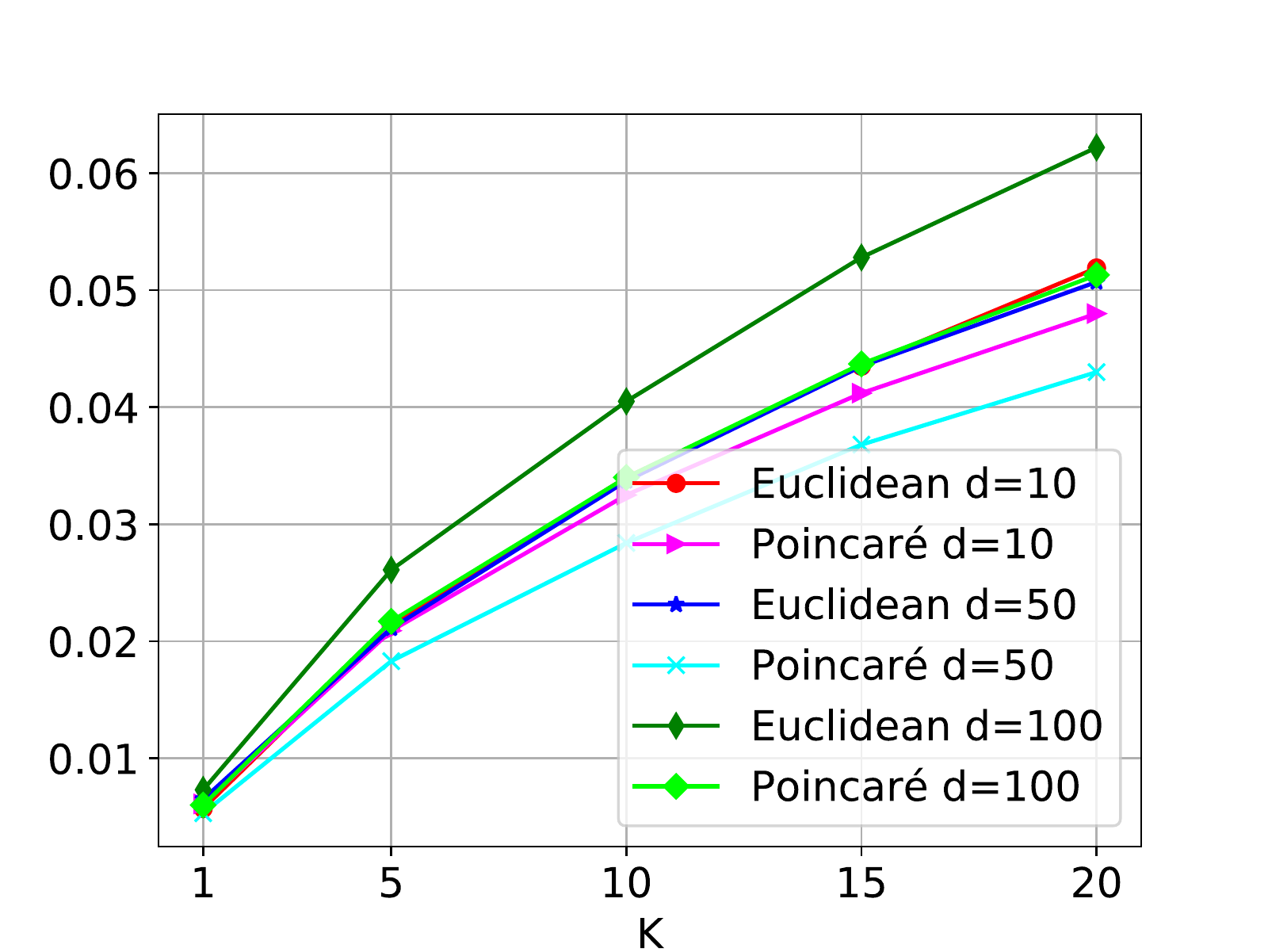}\label{fig:BPR g}}
	\subfloat[Cellphones NDCG]{\includegraphics[width=0.25\linewidth]{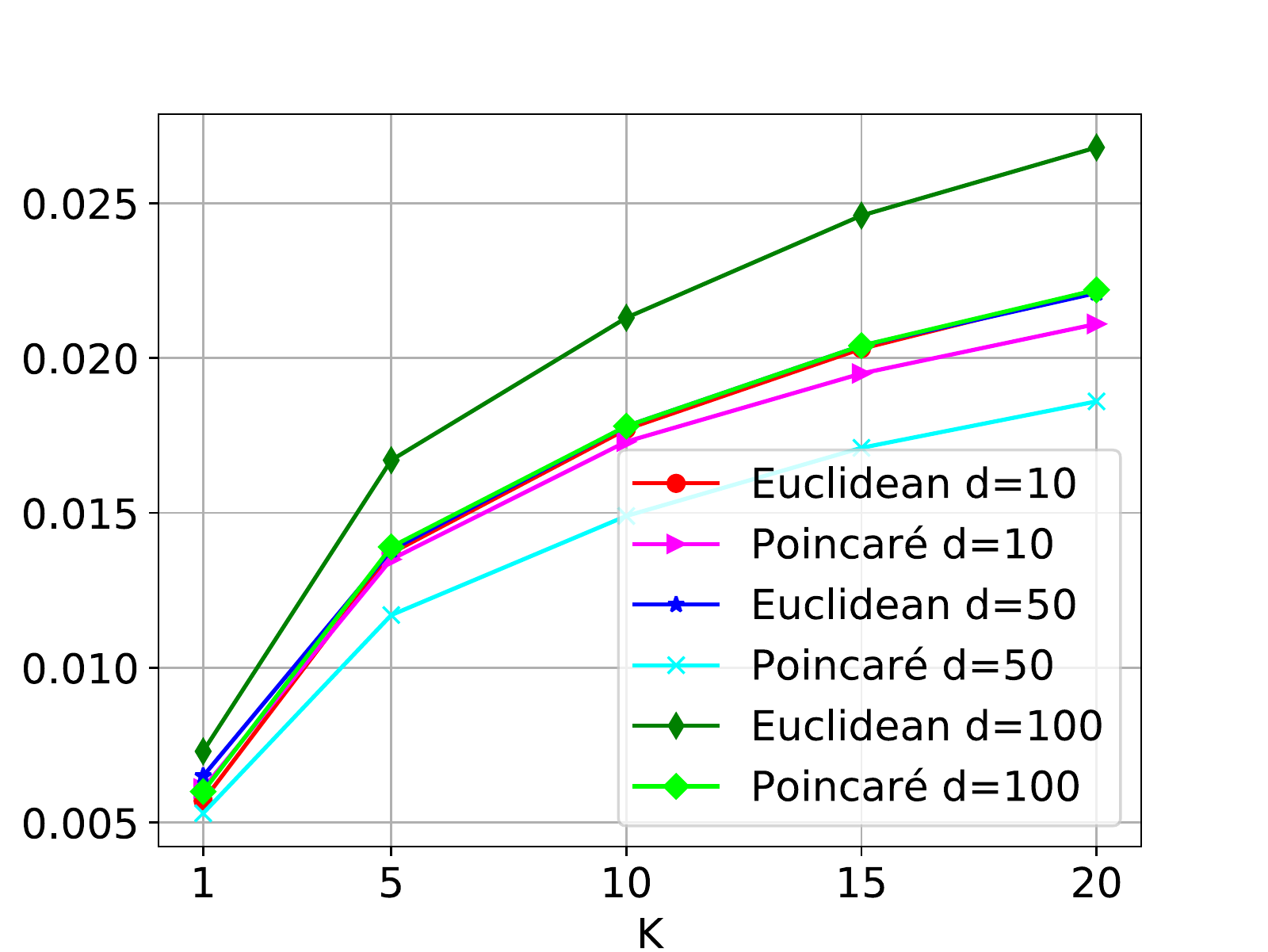}\label{fig:BPR h}}
\vspace{-1.3em}
\caption{MF-BPR HRs and NDCGs on four datasets.}
\label{fig:BPR}
\vspace{-2em}
\end{figure*}

\begin{figure*}[t]
\centering
    \subfloat[Movielens 1M HR]{\includegraphics[width=0.25\linewidth]{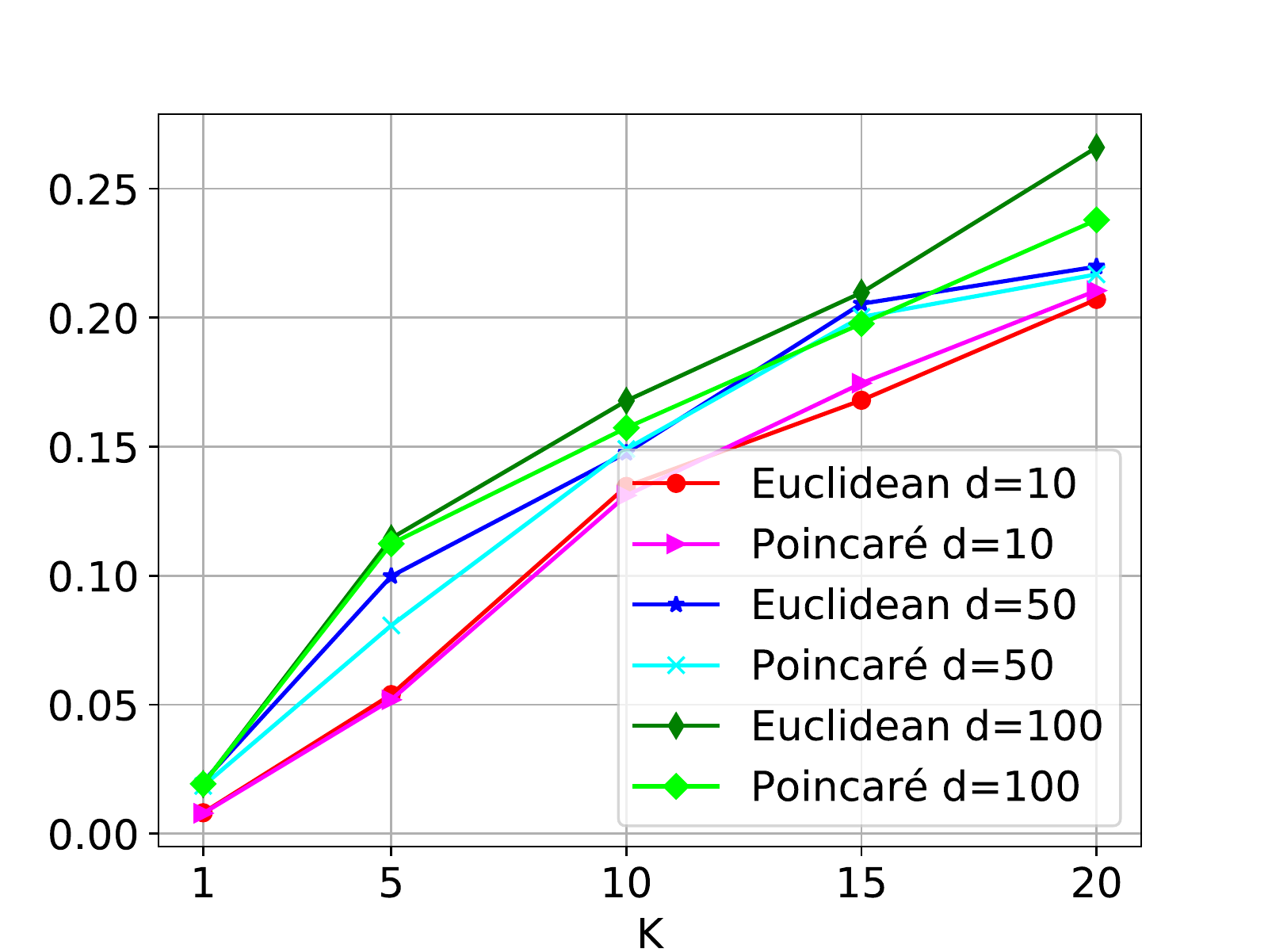}\label{fig:DMF a}}
	\subfloat[Movielens 1M NDCG]{\includegraphics[width=0.25\linewidth]{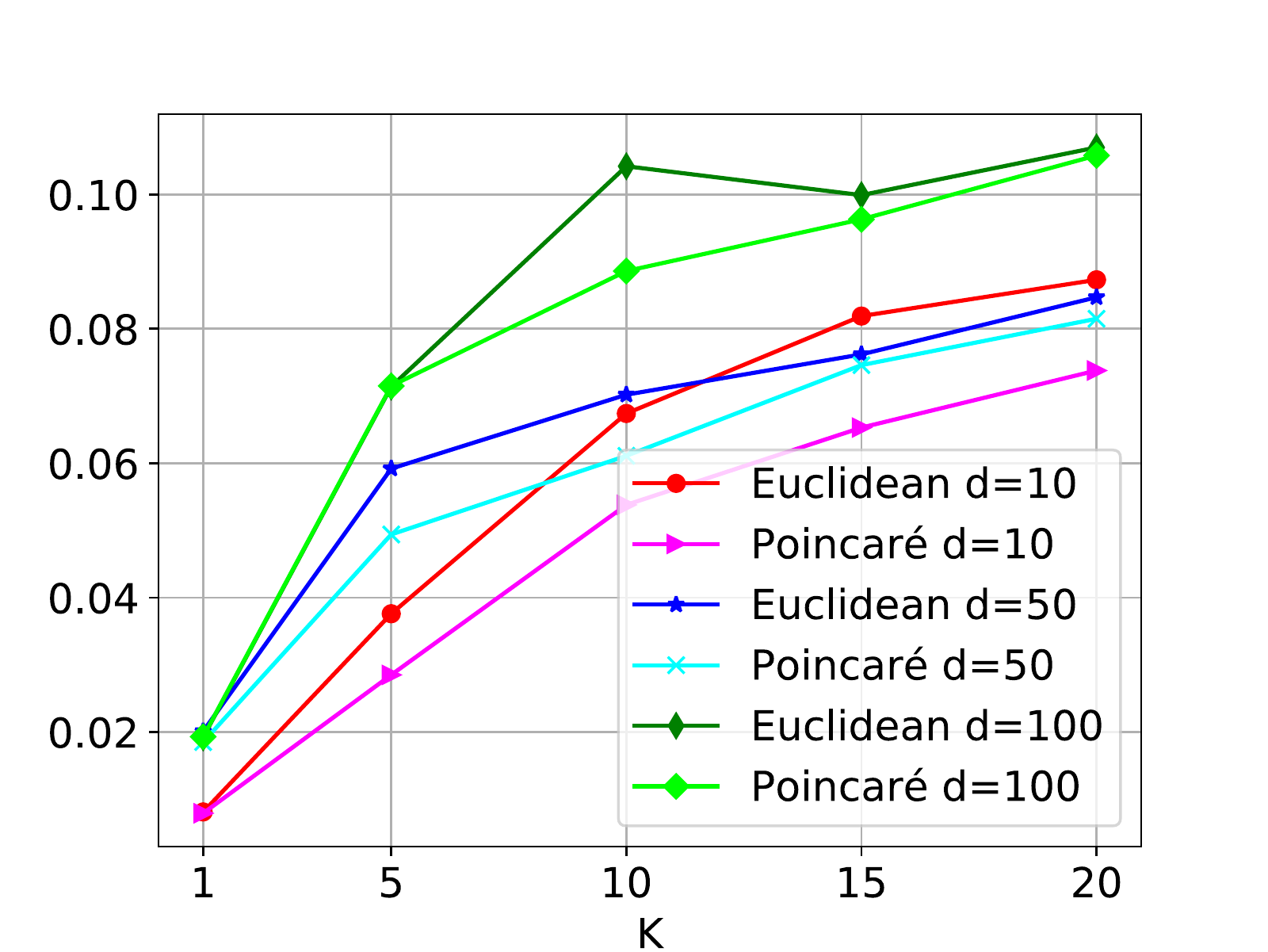}\label{fig:DMF b}}
	\subfloat[Movielens 100K HR]{\includegraphics[width=0.25\linewidth]{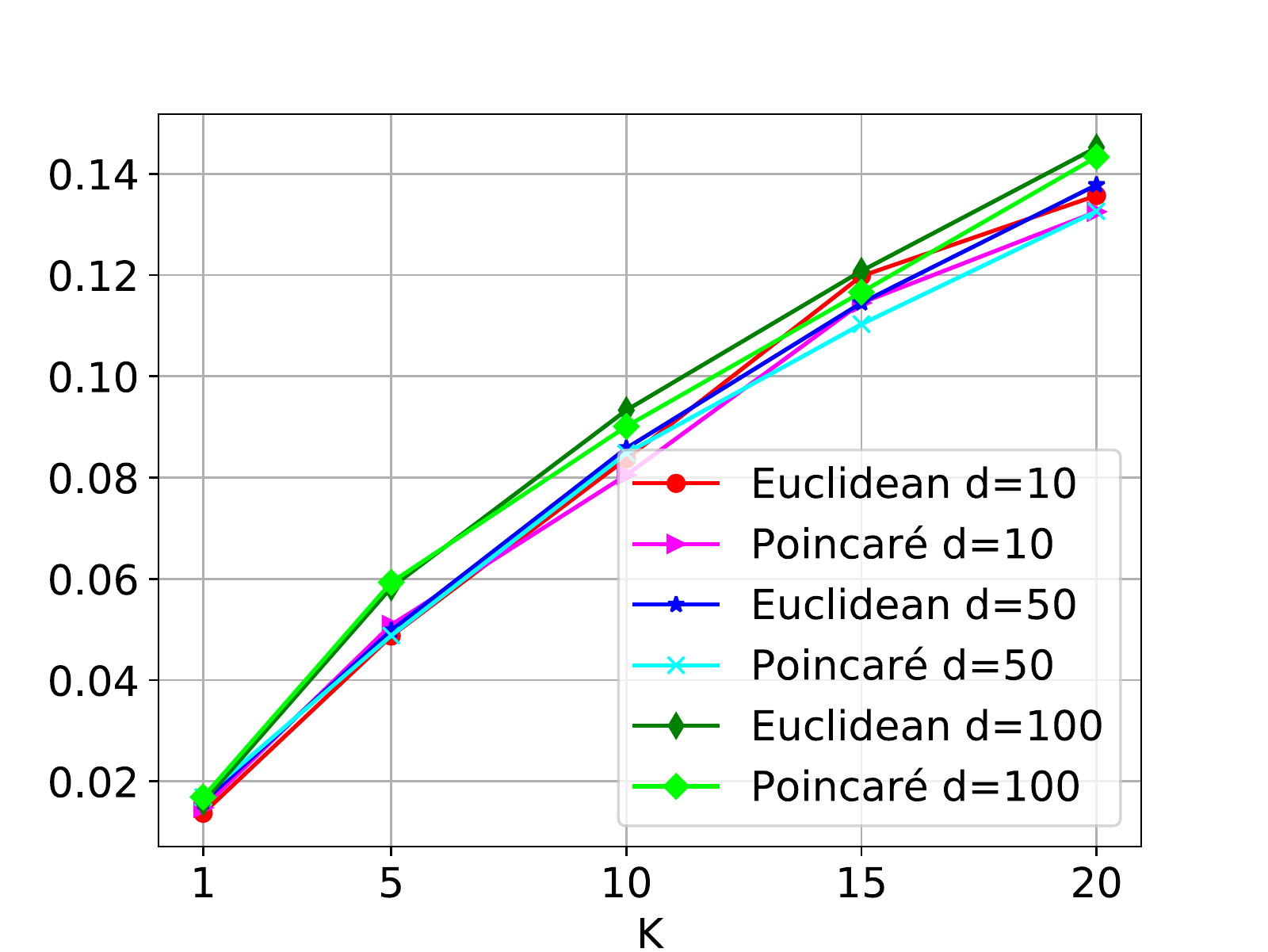}\label{fig:DMF c}}
	\subfloat[Movielens 100K NDCG]{\includegraphics[width=0.25\linewidth]{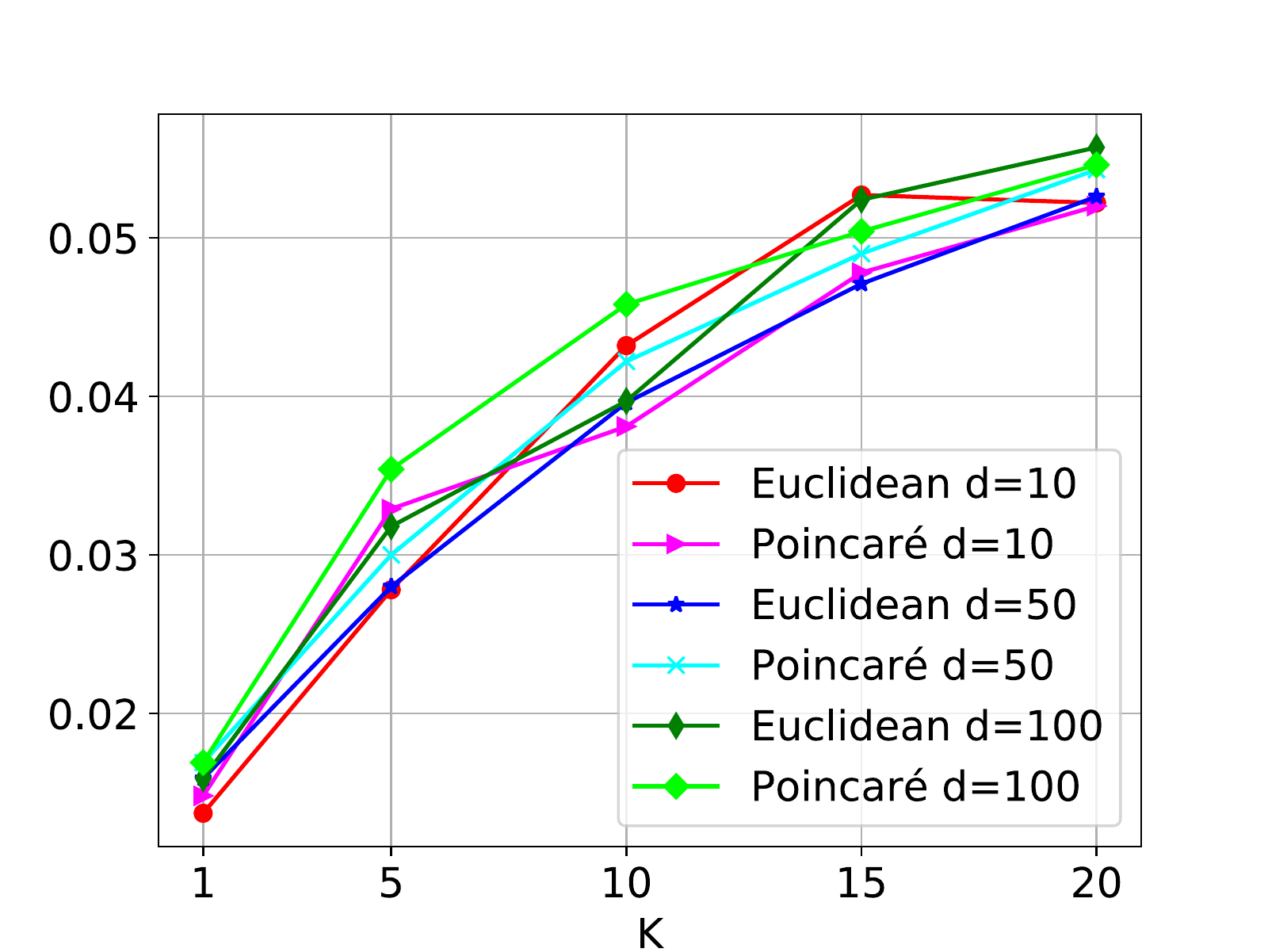}\label{fig:DMF d}}\\
	\vspace{-1.3em}
	\subfloat[Automotive HR]{\includegraphics[width=0.25\linewidth]{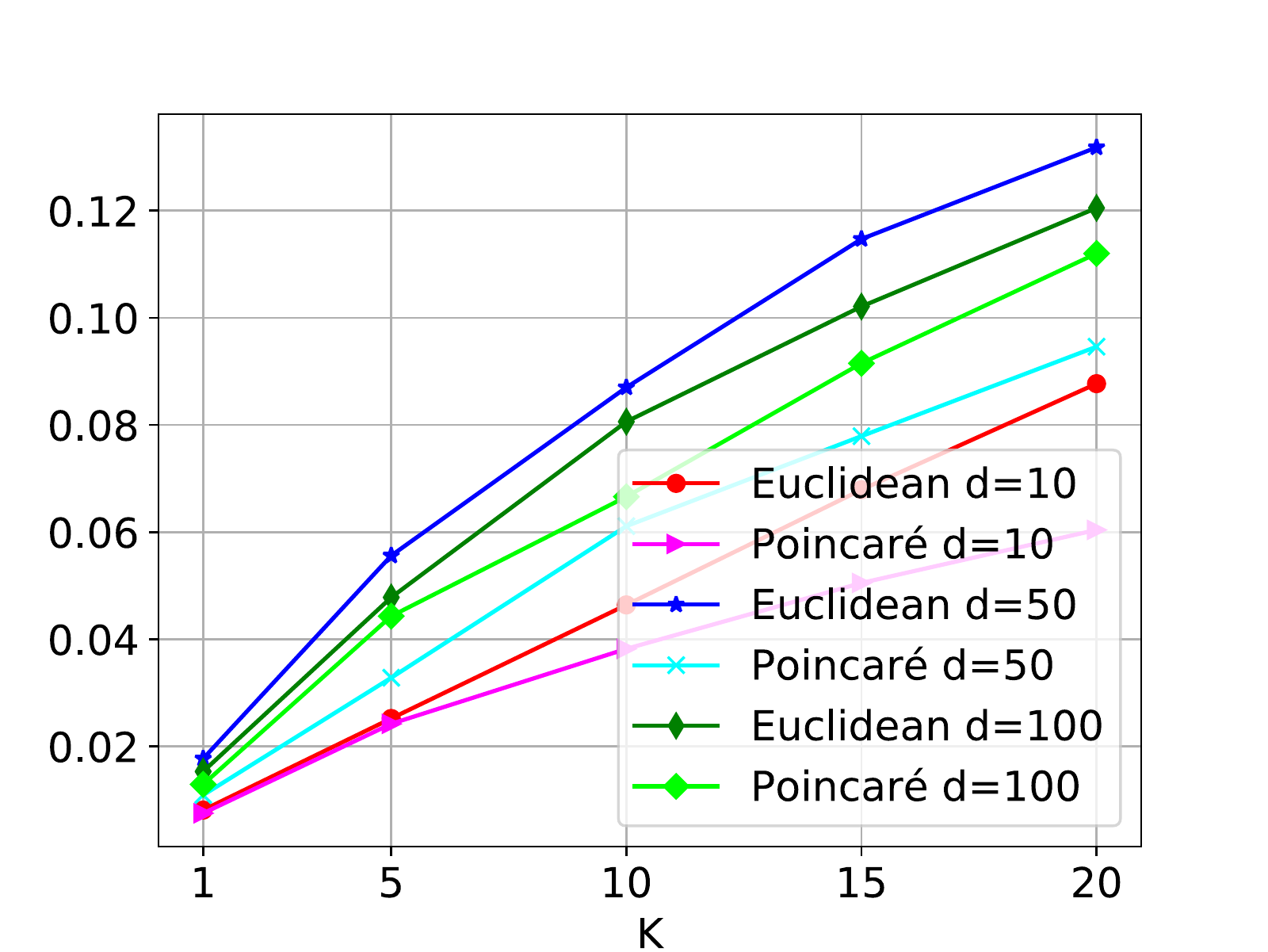}\label{fig:DMF e}}
	\subfloat[Automotive NDCG]{\includegraphics[width=0.25\linewidth]{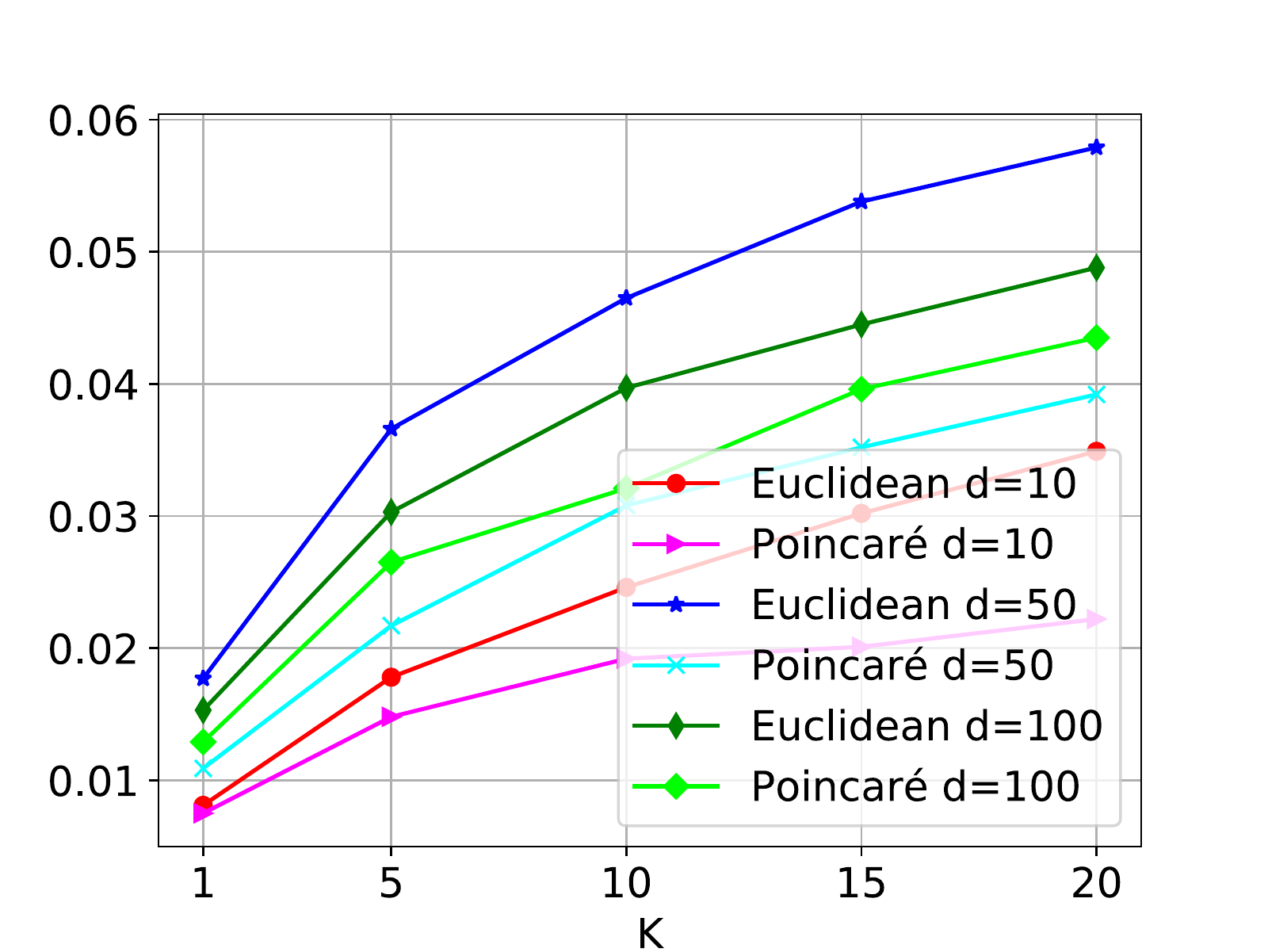}\label{fig:DMF f}}
	\subfloat[Cellphones HR]{\includegraphics[width=0.25\linewidth]{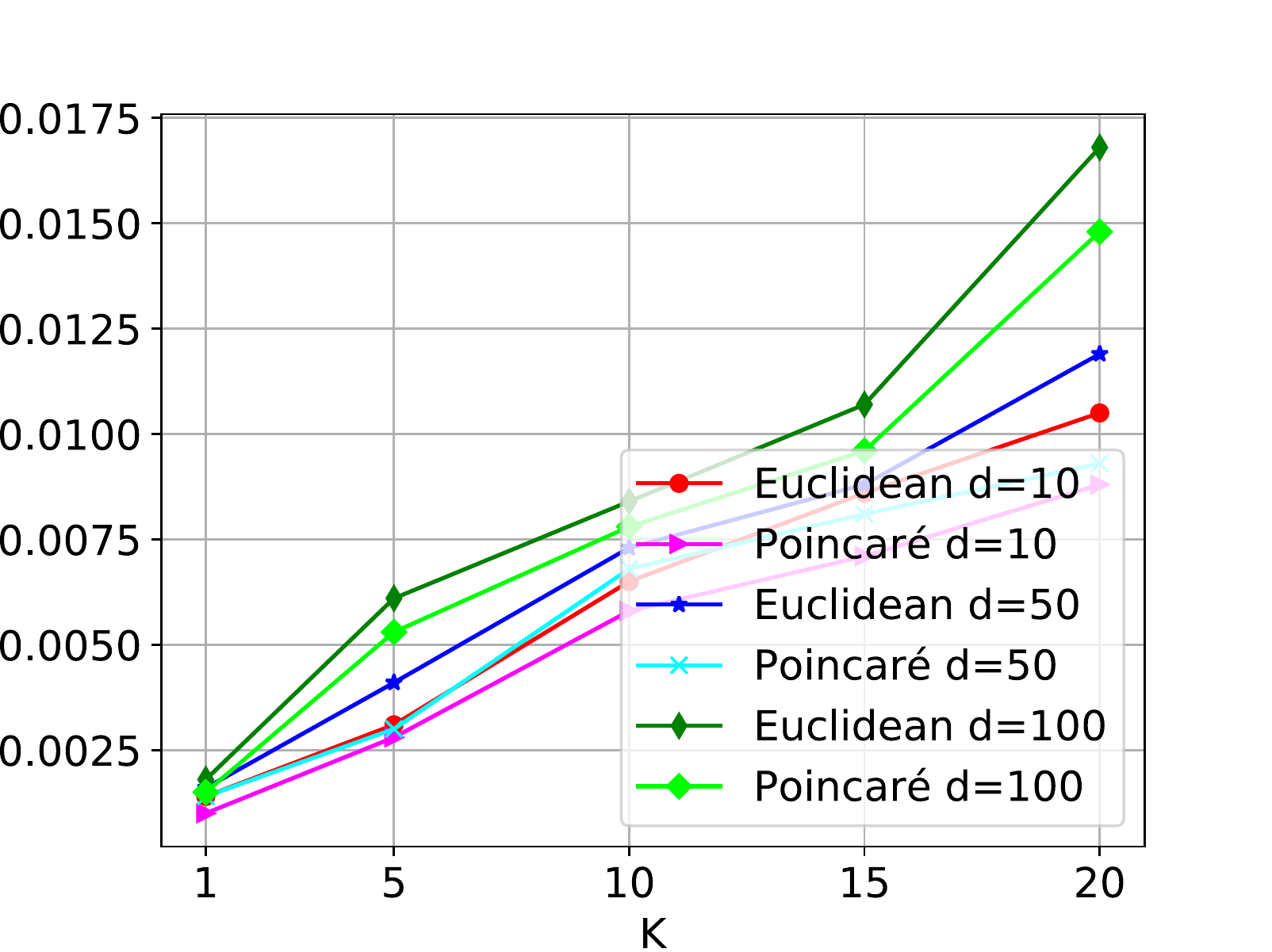}\label{fig:DMF g}}
	\subfloat[Cellphones NDCG]{\includegraphics[width=0.25\linewidth]{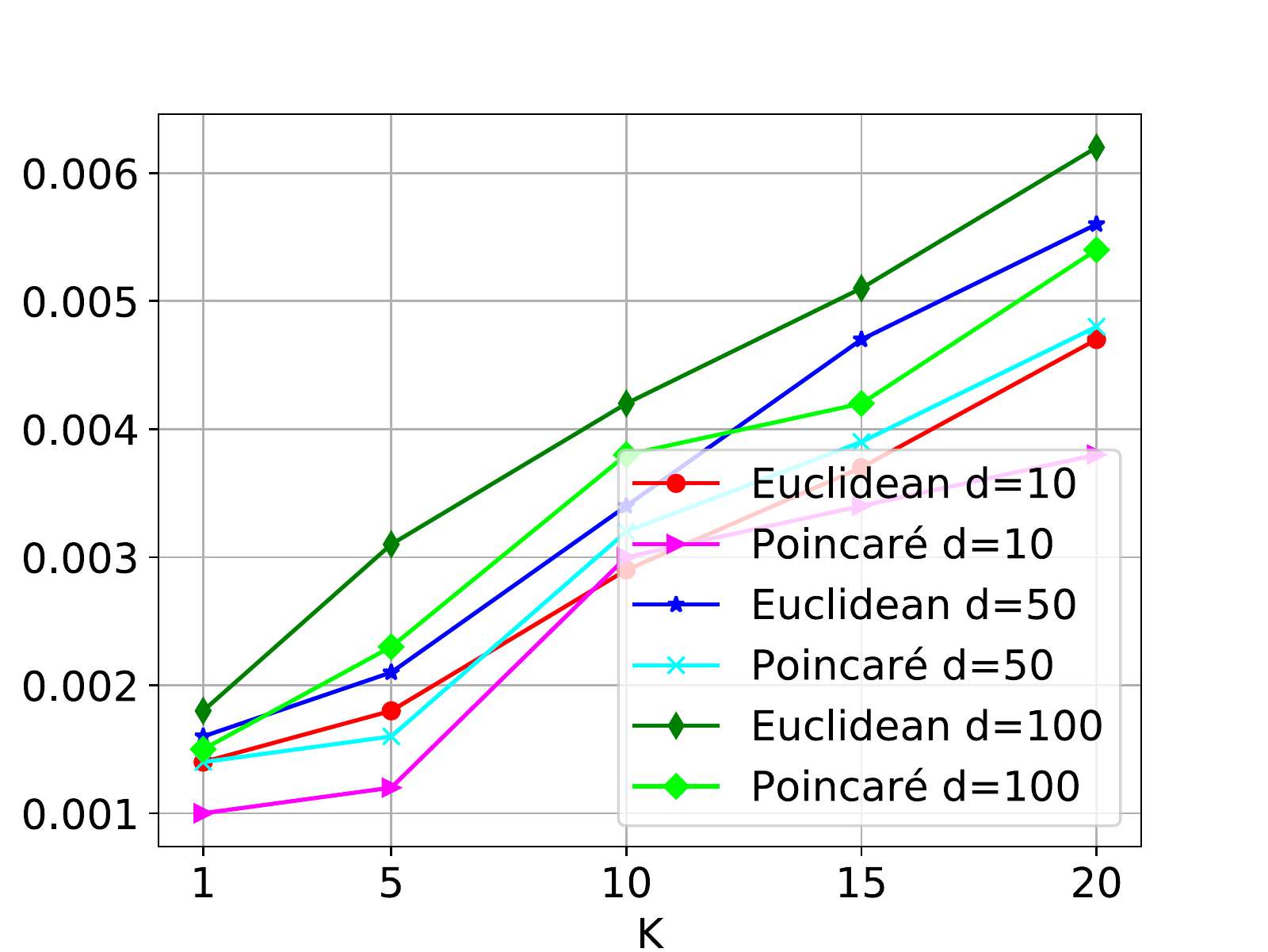}\label{fig:DMF h}}
\vspace{-1.3em}
\caption{DMF HRs and NDCGs on four datasets.}
\label{fig:DMF}
\vspace{-2.5em}
\end{figure*}

\begin{figure*}[t]
\centering
	\subfloat[SoRec on Epinions]{\includegraphics[width=0.25\linewidth]{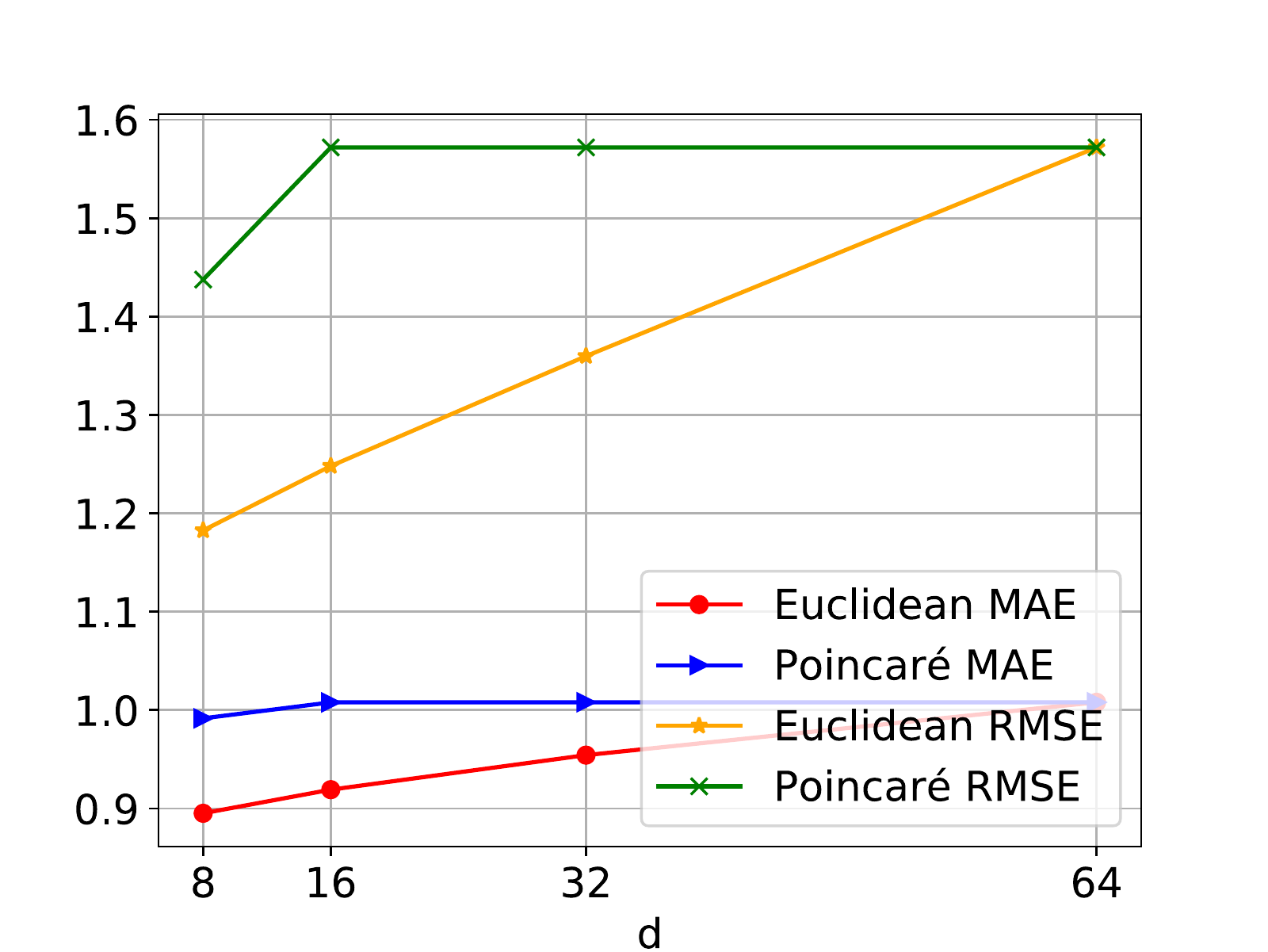}}
	\subfloat[SoRec on Ciao]{\includegraphics[width=0.25\linewidth]{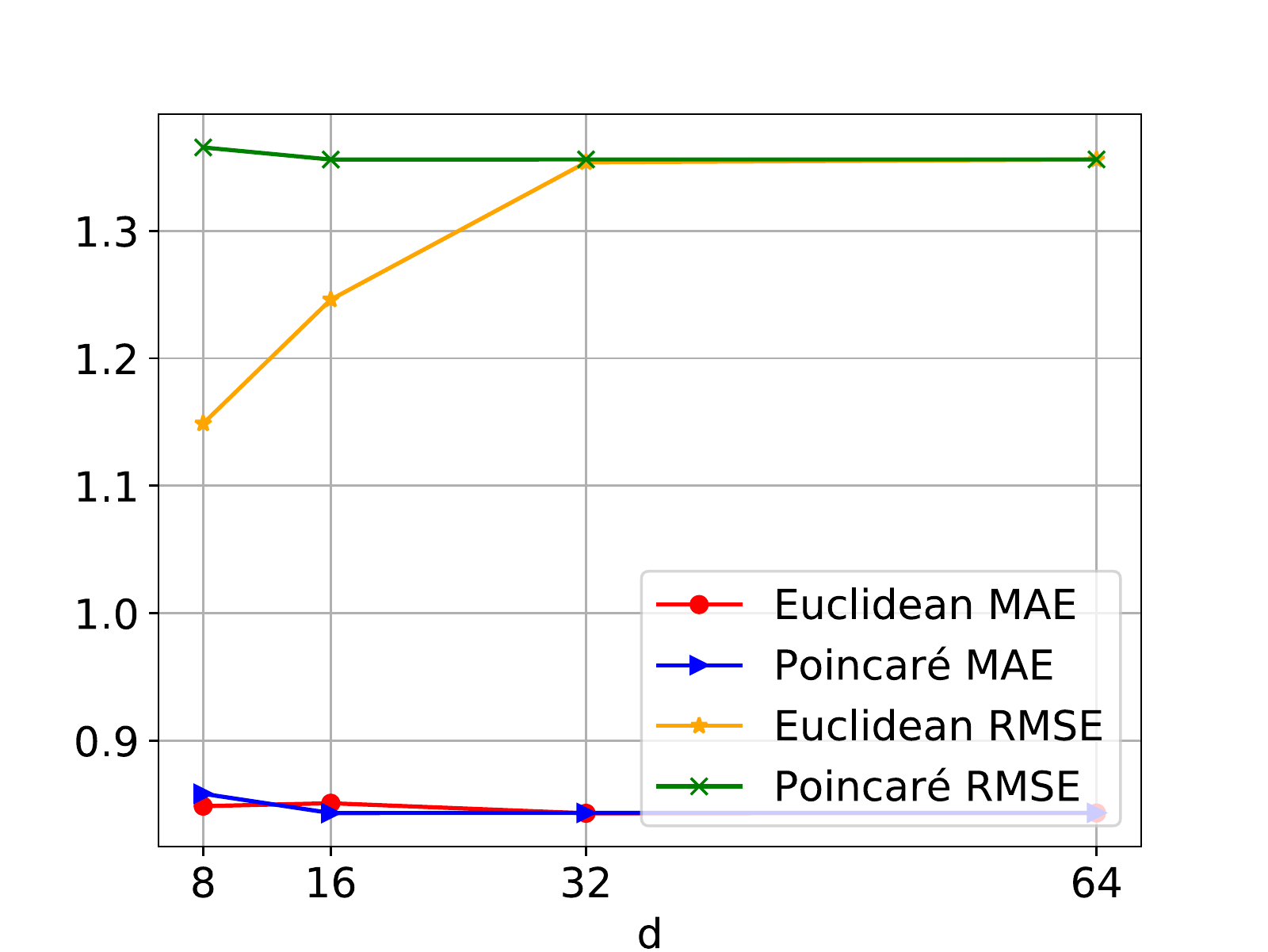}}
	\subfloat[TrustSVD on Epinions]{\includegraphics[width=0.25\linewidth]{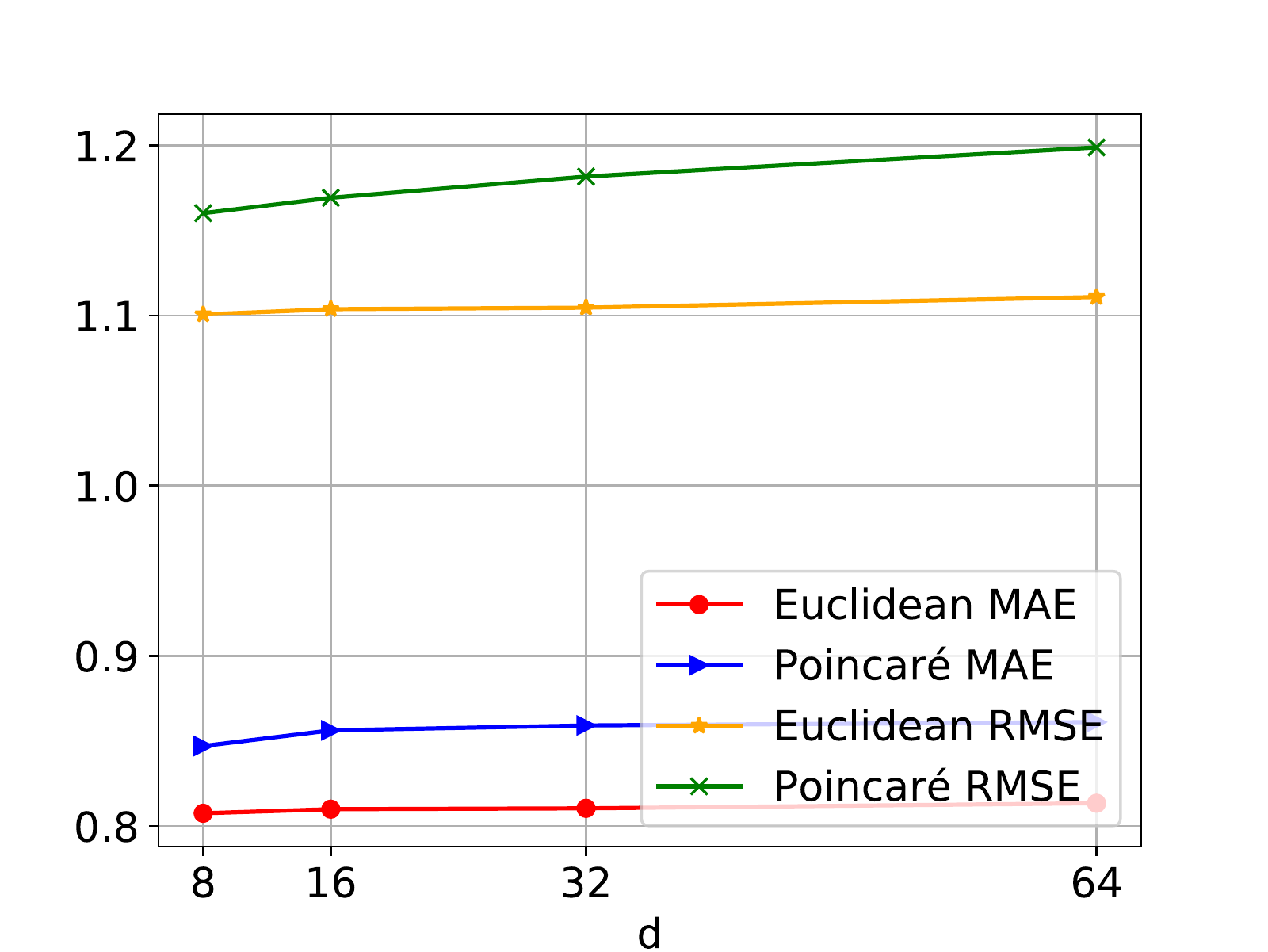}}
	\subfloat[TrustSVD on Ciao]{\includegraphics[width=0.25\linewidth]{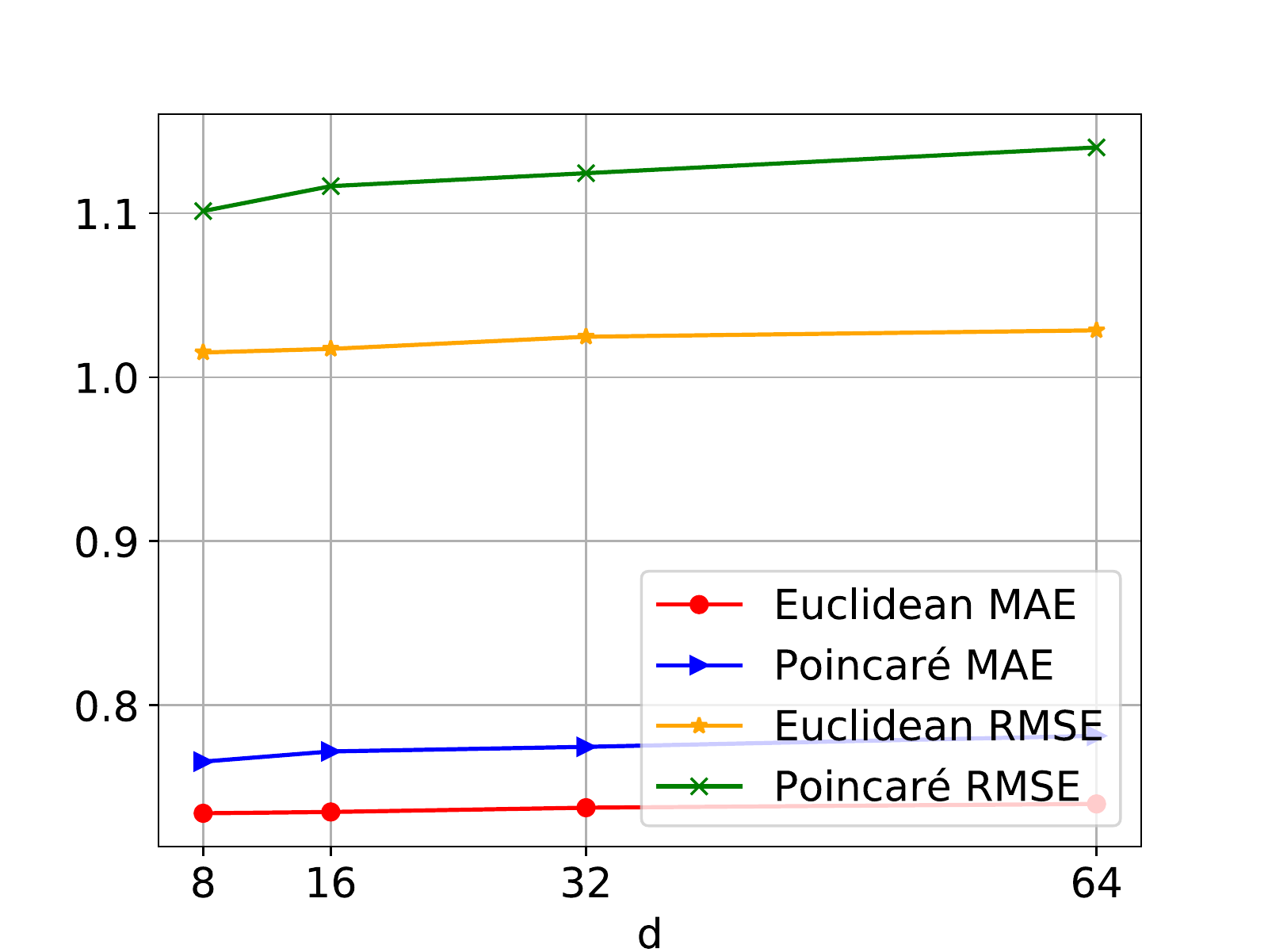}}
\vspace{-1.3em}
\caption{SoRec and TrustSVD MAEs and RMSEs on Epinions and Ciao.}
\label{fig:social recommendation results}
\vspace{-2em}
\end{figure*}

\subsection{\textbf{Evaluation Metrics}}
Some baseline approaches are originally proposed for the top-n recommendation task, others are for the rating prediction task. We keep their tasks unchanged for the best performance. We report the hit ratio (HR) at $k=\{1,5,10,15,20\}$ and normalized discounted cumulative gain (NDCG) at $k=\{1,5,10,15,20\}$ for top-n recommendation tasks, and report the mean absolute error (MAE) and rooted mean square error (RMSE) for rating prediction tasks. 

For rating prediction tasks, we leave 20\% of the ratings as the validation set, and 20\% as the test set, the remaining 60\% as the training set. For top-n ranking tasks, we adopt the popular leave-one-out method. We leave the last item each user has interacted as the test set, and leave the penultimate as the validation set. \textbf{Note that we do not do negative sampling for both HR and NDCG}, instead we rank all the negative items, in this way we can have a more accurate and convincing result and avoid the bias introduced by the sampling \cite{krichene2020sampled}.

\textbf{Hit Ratio.} For a $\it{single}$ user, leave one item out from the user's list of interacted items (usually the last one that the user has interacted with). Then rank the left out positive item together with all the negative items that the user has not interacted with. If the left out item is at the top $n$ of the ranking list, then we consider this item as a hit. The hit ratio of the recommender system is then the total number of hits divided by the number of users.

\textbf{Normalized Discounted Cumulative Gain.} We first explain what cumulative gain (CG) and discounted cumulative gain (DCG) are. Given a list of top $n$ recommendations, $r_{i}=1$ if the $i$-th item has been interacted or is liked by the user; $r_{i}=0$ if the $i$-th item has not been interacted or is disliked by the user. Thus, CG and DCG are calculated by
\begin{equation}
\small
    CG@n=\sum_{i=1}^{n}r_{i},\ \  DCG@n=\sum_{i=1}^{n}r_{i}\cdot\frac{1}{\text{log}_{2}(i+1)}
\end{equation}
CG measures the relevance of the recommendation list. DCG improves it by taking into consideration the ordering of the items. But DCG is not normalized so it may vary for different users. Therefore, we use the ideal discounted cumulative gain (IDCG) as the normalization constant, which is calculated by moving all the interacted items to the top of the list. This is an $ideal$ list and it has the largest DCG value. Therefore, NDCG is always between 0 and 1:
\begin{equation}
\small
\label{eq: ndcg}
    NDCG@n=\frac{DCG@n}{IDCG@n}
\end{equation}

\textbf{Mean Absolute Error and Rooted Mean Square Error.} MAE and RMSE are measures of errors between predicted ratings and observed ratings. They is calculated using the following equations:
\begin{equation}
\small
    MAE=\frac{\sum_{i=1}^{n}|y_{i}-x_{i}|}{n},\ \ RMSE=\sqrt{\frac{\sum_{i=1}^{n}(y_{i}-x_{i})^{2}}{n}}
\end{equation}
where $y$ is the observed rating and $x$ is the predicted rating.

\subsubsection{\textbf{Experiment Settings}}
Note that the target of our experiment is to compare the performance of the Euclidean space and the hyperbolic space on various latent space models, not to compare the performance of different models. For each model, the Euclidean setting and the hyperbolic setting share some basic parameters like the learning rate and batch size, we keep such parameters to be the same in both Euclidean setting and hyperbolic setting. Other latent space specified parameters such as the margin in CML is tuned individually in Euclidean setting and hyperbolic setting to reach their best performance.

We report the test scores based on the best validation scores. For each experiment, we do five independent trials and report the average result. For all methods, the learning rate is tuned among $\{0.1,0.01,0.001,0.0001\}$ using Adam optimizer. The batch size of each dataset is tuned among $\{50,500,5000,50000\}$ for a relatively good performance and short running time. We set hyperbolic curvature to be constant -1. For top-n recommendation tasks, we report HRs and NDCGs when the latent size $d=\{10,50,100\}$. For rating prediction tasks, we report MAE and RMSE when the latent size $d=\{8,16,32,64\}$. For CML, the margin is tuned from 0.1 to 2.0 for Euclidean space, and from 2 to 40 for hyperbolic space (the norm of the hyperbolic embedding is clipped by approximately 6 due to the precision). We keep other hyper-parameters as suggested by the original papers.

\subsubsection{\textbf{Results}}
The results of our experiments are shown from Fig. 2 to Fig. 5. Following the hypotheses we made in \autoref{sec:hypotheses}, we first compare the performance of distance models and projection models, then compare different datasets, last we compare different latent sizes.

\textbf{Distance Models vs. Projection Models.} According to the results, hyperbolic space outperforms Euclidean space in distance models such as CML. For example, from the results of CML on Automotive and Cellphones shown in \autoref{fig:CML e}, \autoref{fig:CML f}, \autoref{fig:CML g}, and \autoref{fig:CML h}, the numbers of hyperbolic space have a great improvement over the numbers of Euclidean space. One interesting thing is that for Movielens shown in \autoref{fig:CML a}, \autoref{fig:CML b}, \autoref{fig:CML c}, and \autoref{fig:CML d}, hyperbolic space and Euclidean space seem to have a similar performance. This will be explained when we compare different datasets in the next part. For other projection models including MF-BPR in \autoref{fig:BPR}, DMF in \autoref{fig:DMF}, TrustSVD and SoRec in \autoref{fig:social recommendation results}, hyperbolic space can't outperform Euclidean space in any of them. Therefore, we can draw a conclusion that distance models are more suited for hyperbolic space than projection models.

\textbf{High Density Datasets vs. Low Density Datasets.} The performance of Euclidean space and hyperbolic space on Movielens 1M and Movielens 100K are much closer than Automotive and Cellphones for general item recommendation methods as shown in \autoref{fig:CML}, \autoref{fig:BPR}, and \autoref{fig:DMF}. In \autoref{fig:CML}, the performance of Euclidean space even becomes comparable with hyperbolic space on Movielens. From \autoref{table:datasets statistics}, we can see that Movielens 1M and Movielens 100K have a very high density compared with other datasets. Based on the above observations and our analysis in \autoref{sec:hypotheses}, we can draw a conclusion that datasets with a lower density can benefit more from hyperbolic space.

\textbf{Influence of the Latent Size.} From the results of CML on Automotive and Cellphones as shown in \autoref{fig:CML e}, \autoref{fig:CML f}, \autoref{fig:CML g}, and \autoref{fig:CML h}, we can see that the improvement of hyperbolic space over Euclidean space decreases as the latent size increases. This serves as an evidence of our hypothesis about the latent size made in \autoref{sec:hypotheses}. We can draw a conclusion that hyperbolic space is better than Euclidean space in metric learning based approaches especially when the latent size is small, but when the latent size is large enough, Euclidean space becomes comparable with hyperbolic space.
\vspace{-1em}

\subsubsection{\textbf{Drawbacks of Hyperbolic Space}}
Although hyperbolic space can outperform Euclidean space, there are still some drawbacks when using hyperbolic space.

\textbf{High Computational Complexity.} When using hyperbolic space, the exponential map and logarithmic map are necessary whenever we apply a simple operation to the embedding such as matrix multiplication and bias addition. This makes the algorithm computational inefficient. It does not cause much trouble in some straightforward methods which do not have much computational work such as CML and SoRec, but greatly increases the computational resources needed in some complicated approaches such as DMF which involves neural network structures. 

\textbf{Pay Attention to Invalid Values.} Since the computation involves inverse hyperbolic functions, it is extremely easy for the denominators to be 0 or infinite. Due to the precision of the device, it is necessary to clip values wherever might become invalid.

\subsubsection{\textbf{Comments on Using Hyperbolic Space}}
Based on the above results, we can now make several comments and suggestions for using hyperbolic space and learning hyperbolic embeddings.
\begin{itemize}[leftmargin=1em]
    \item Distance models are more suited for learning hyperbolic embeddings than projection models.
    \item If the density of the dataset is large, Euclidean space should be a better choice because it has a comparable performance with hyperbolic space and the computational complexity is low; when the density is low, hyperbolic space is more preferable because it will have a much better performance than Euclidean space.
    \item Choose an appropriate latent size. In most cases, hyperbolic embeddings only need a relatively small latent size to achieve good performance, which can help to save resources.
\end{itemize}

%% file: methodology.tex
\begin{figure*}[h]
\centering
    \subfloat[Epinions HR]{\includegraphics[width=0.25\linewidth]{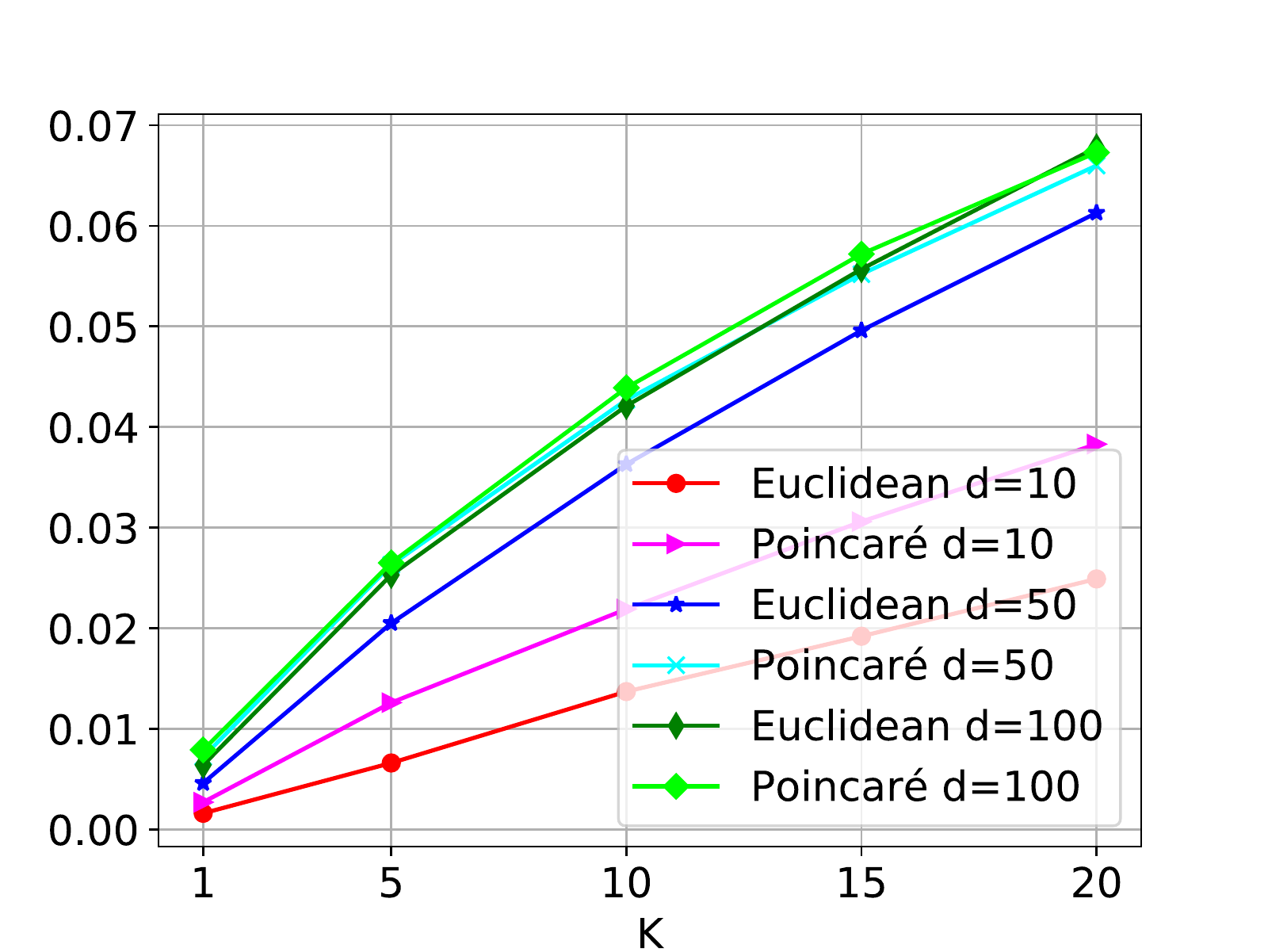}\label{fig:SCML a}}
	\subfloat[Epinions NDCG]{\includegraphics[width=0.25\linewidth]{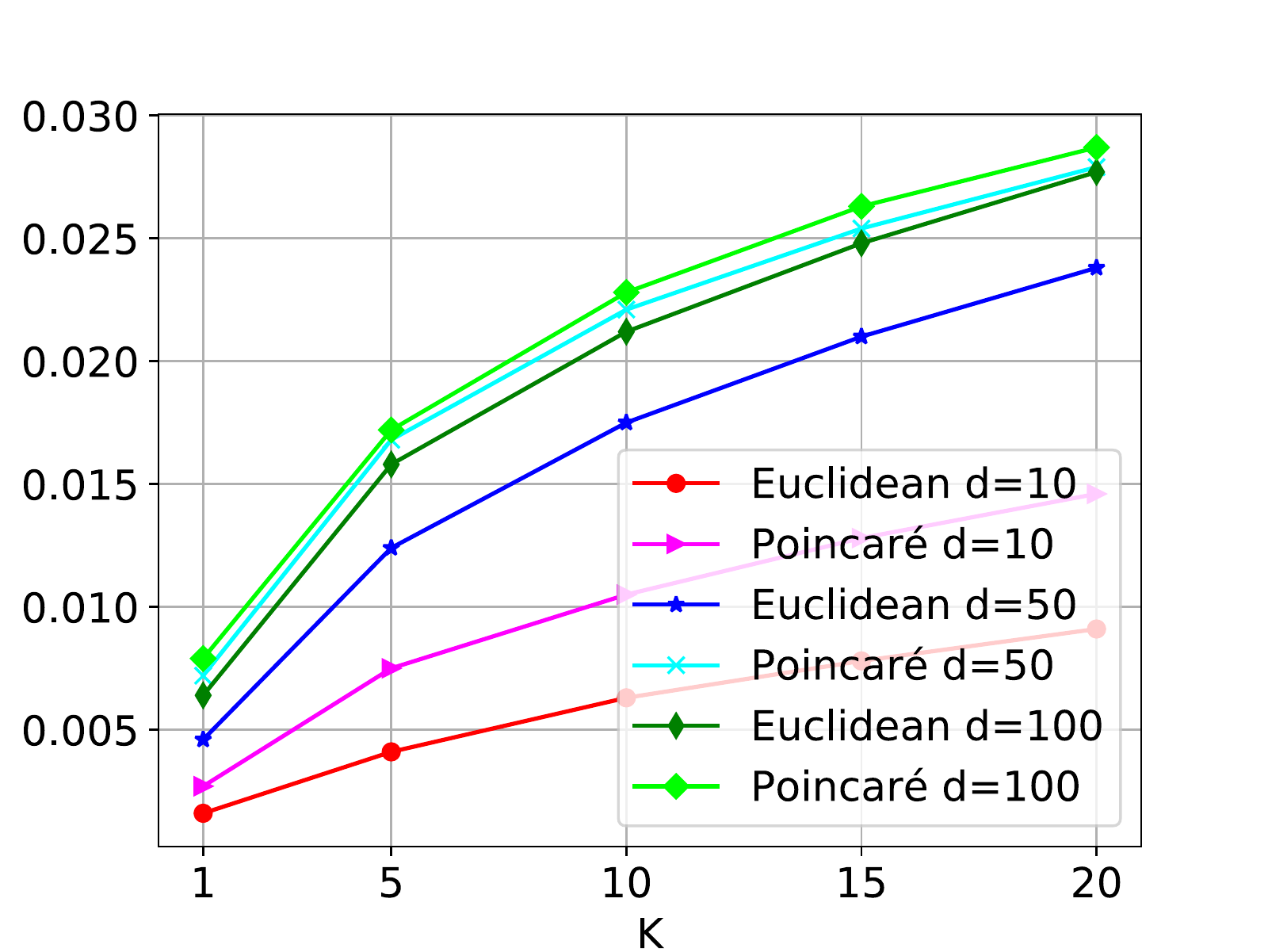}\label{fig:SCML b}}
	\subfloat[Ciao HR]{\includegraphics[width=0.25\linewidth]{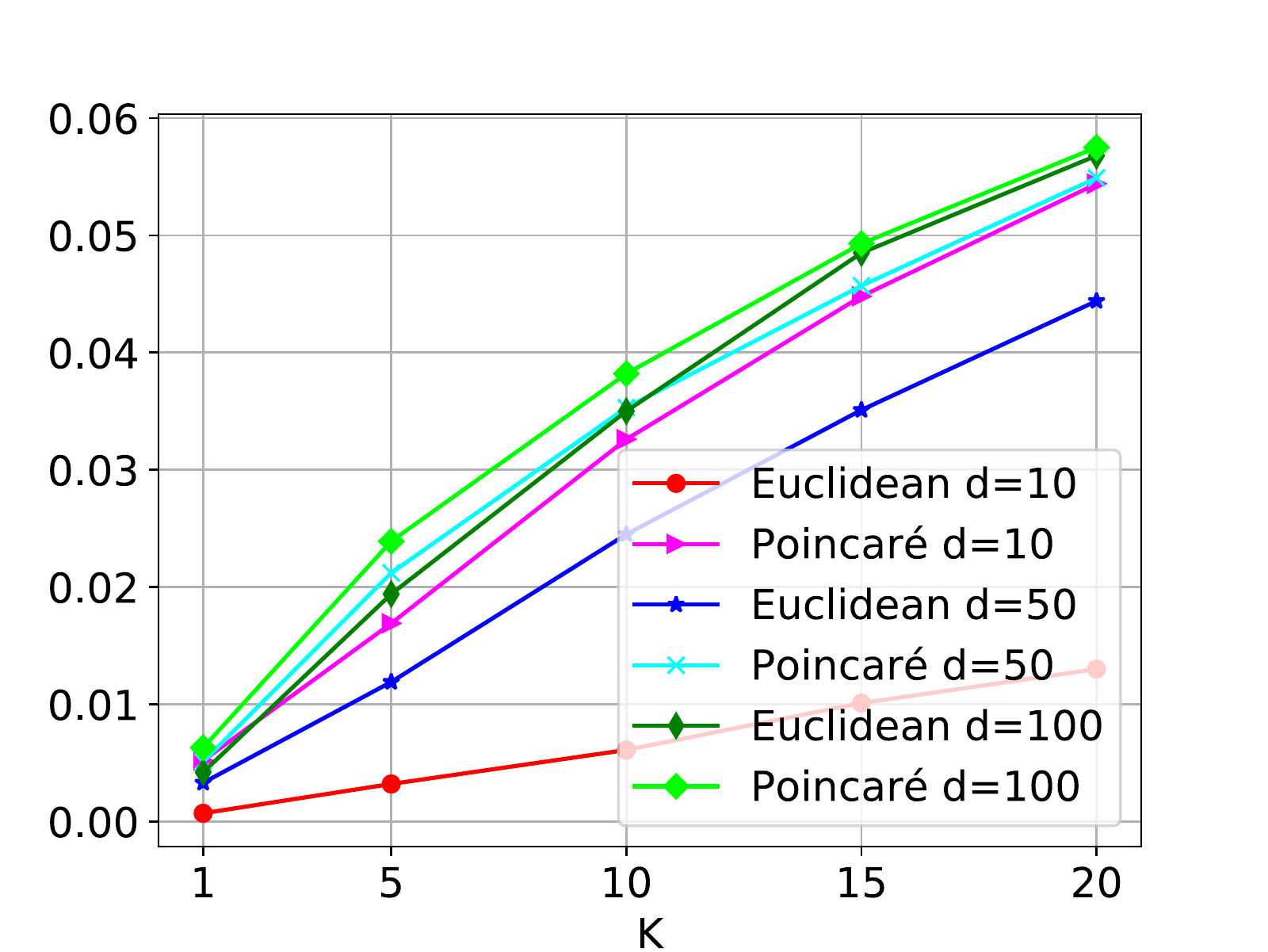}\label{fig:SCML c}}
	\subfloat[Ciao NDCG]{\includegraphics[width=0.25\linewidth]{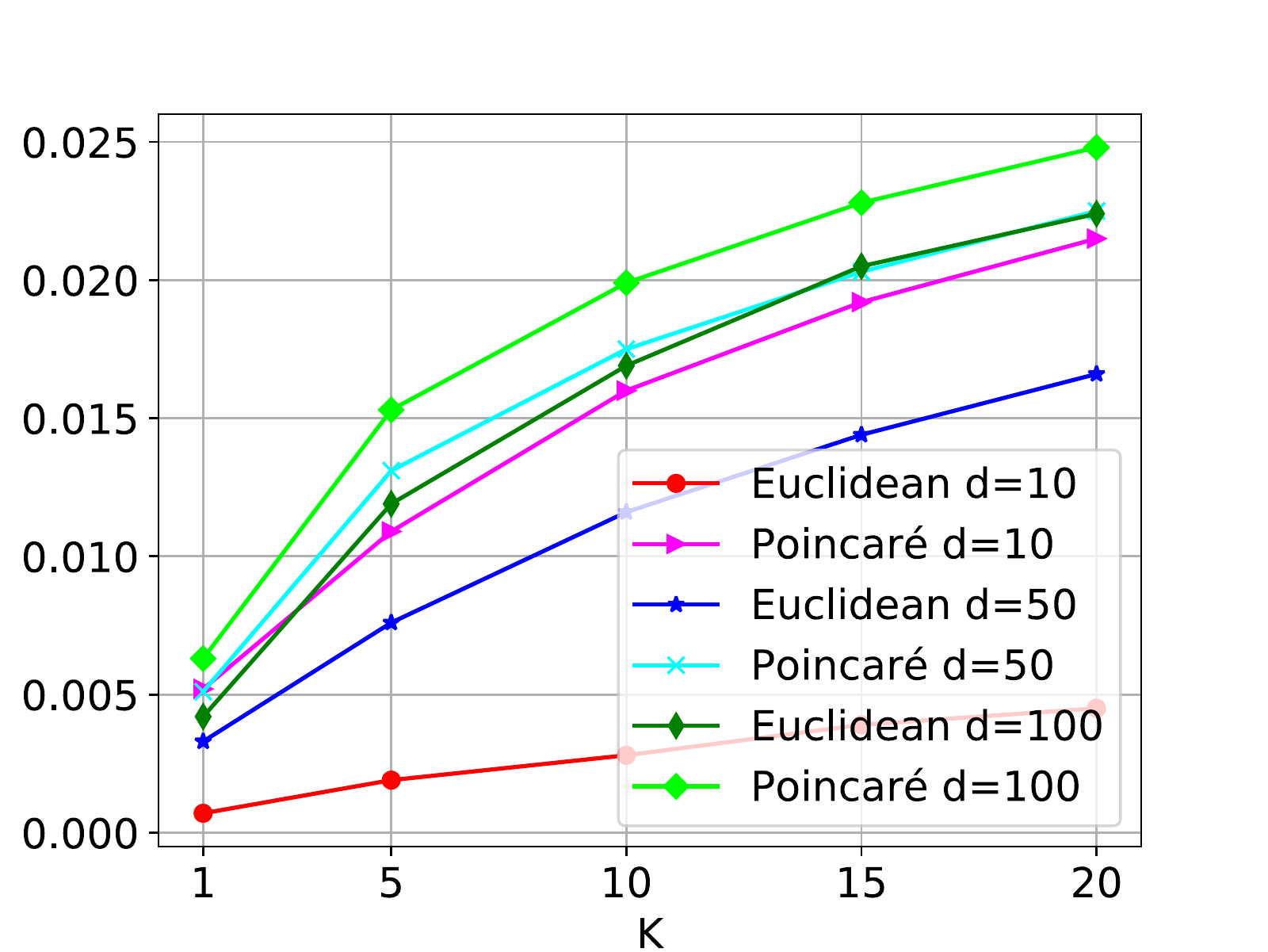}\label{fig:SCML d}}
\vspace{-1.3em}
\caption{SCML HRs and NDCGs on Epinions and Ciao.}
\label{fig:SCML}
\vspace{-1.3em}
\end{figure*}

\section{Social Collaborative Metric Learning}

As we explained in \autoref{sec:baselines} that there is a lack in using distance models to solve top-n social recommendation task, we here propose a new metric learning based social recommendation method for top-n recommendation called Social Collaborative Metric Learning (SCML), which is a generalized form of Collaborative Metric Learning (CML) \cite{hsieh2017collaborative}. We apply SCML in both Euclidean space and hyperbolic space using their corresponding distance functions. We call hyperbolic SCML as HSCML.

Suppose the set of user-item positive pairs is $\mathcal{R}$, and the set of user-user positive pairs is $\mathcal{S}$. For user $i$, sample a positive item $j$ and a negative item $k$ such that $(i,j)\in\mathcal{R}, (i,k)\notin\mathcal{R}$, and sample a positive neighbor $m$ and a negative neighbor $n$ such that $(i,m)\in\mathcal{S}, (i,n)\notin\mathcal{S}$. The item side loss $\mathcal{L}_{item}$ and the social connection side loss $\mathcal{L}_{so}$ are defined as
\begin{equation}
\small
\label{eq:item loss}
    \mathcal{L}_{item}=\sum_{(i,j)\in\mathcal{R}}\sum_{(i,k)\notin\mathcal{R}}[m_{item}+d(i,j)^{2}-d(i,k)^{2}]_{+}
\end{equation}
\begin{equation}
\small
\label{eq:social loss}
    \mathcal{L}_{so}=\sum_{(i,m)\in\mathcal{S}}\sum_{(i,n)\notin\mathcal{S}}[m_{so}+d(i,m)^{2}-d(i,n)^{2}]_{+}
\end{equation}
where $m_{item}$ and $m_{so}$ are the safety margins, $[\cdot]_{+}=max(\cdot,0)$ is the standard hinge loss. $d(\cdot, \cdot)$ is the Euclidean distance function in SCML, and is the hyperbolic distance function in HSCML. The loss function of our model is
\begin{equation}
\small
    \mathcal{L}=\mathcal{L}_{item}+\lambda\mathcal{L}_{so}
\end{equation}
where $\lambda$ controls the weight of $\mathcal{L}_{so}$. We use $\mathcal{L}_{item}$ to push the positive user-item pairs close to each other and push the negative user-item pairs away to each other, and use $\mathcal{L}_{so}$ to push the socially related users close to each other. Thus, for a particular user, the items that have been interacted by his or her socially connected neighbors are also being pushed close to the user. This reflect the fact that users are more likely to interact with the items that their friends have interacted with.

\begin{table*}[h]
\begin{center}
 \begin{tabular}{|c|c|c|c|c|c|c|c|c|c|c|c|} 
 \hline
 Dataset & Metric & BPR & FISM & NeuMF & CML & LRML & SBPR & CUNE-BPR & RML-DGATs & SAMN & HSCML\\
 \hline
 \multirow{4}{*}{Ciao} & HR@10 & 0.2286 & 0.2076 & 0.2328 & 0.2302 & 0.2471 & 0.2364 & 0.2823 & \textbf{0.3108} & \underline{0.3098} & 0.3052\\
 \cline{2-12}
 & NDCG@10 & 0.1379 & 0.1207 & 0.1385 & 0.1365 & 0.1473 & 0.1446 & 0.1704 & 0.1803 & \textbf{0.2049} & \underline{0.2018}\\
 \cline{2-12}
 & HR@20 & 0.3074 & 0.2847 & 0.3147 & 0.3129 & 0.3184 & 0.3282 & 0.3729 & \textbf{0.4157} & \underline{0.3863} & 0.3844\\
 \cline{2-12}
 & NDCG@20 & 0.1579 & 0.1403 & 0.1601 & 0.1563 & 0.1703 & 0.1625 & 0.1861 & 0.2066 & \textbf{0.2242} & \underline{0.2194}\\
 \hline
 \multirow{4}{*}{Epinions} & HR@10 & 0.4104 & 0.3842 & 0.4017 & 0.4118 & 0.4327 & 0.3749 & 0.4377 & \textbf{0.4625} & 0.4422 & \underline{0.4526} \\
 \cline{2-12}
 & NDCG@10 & 0.2591 & 0.2421 & 0.2489 & 0.2562 & 0.2586 & 0.2436 & 0.2598 & 0.2728 & \underline{0.2892} & \textbf{0.2971}\\
 \cline{2-12}
 & HR@20 & 0.5087 & 0.4850 & 0.5023 & 0.5051 & 0.5319 & 0.4868 & \underline{0.5583} & \textbf{0.6018} & 0.5362 & 0.5481\\
 \cline{2-12}
 & NDCG@20 & 0.2839 & 0.2676 & 0.2881 & 0.2972 & 0.3017 & 0.2653 & 0.2806 & 0.3080 & \underline{0.3129} & \textbf{0.3189}\\
 \hline
\end{tabular}
\caption{Top-n recommendation performance comparison. The best performance is in boldface and the second is underlined. }
\label{table:performance comparison}
\end{center}
\vspace{-3em}
\end{table*}

\section{Evaluating SCML and HSCML}
Like what we did in \autoref{sec:evaluatin hyperbolic space} when evaluating the performance of hyperbolic space, we compare the performance of SCML and HSCML. Besides, in order to show how hyperbolic space performs compared with state-of-the-art Euclidean models, we compare the performance of HSCML with Euclidean baseline approaches on Ciao and Epinions using some of the experiment results in the paper of RML-DGATs \cite{wang2020relational}. The baselines include MF-BPR \cite{rendle2012bpr}, FISM \cite{kabbur2013fism}, NeuMF \cite{he2017neural}, CML \cite{hsieh2017collaborative}, LRML \cite{tay2018latent}, SBPR \cite{zhao2014leveraging}, CUNE-BPR \cite{doi:10.1137/1.9781611974973.43}, SAMN \cite{chen2019social} and RML-DGATs \cite{wang2020relational}. We aim to show that, even with a simple hyperbolic distance model such as HSCML, hyperbolic space can still outperform most of the Euclidean baselines and have comparable performance with state-of-the-art methods.


\subsection{Experiment Settings}
First, we compare the performance of Euclidean space and hyperbolic space by comparing SCML and HSCML to evaluate the correctness of our conclusions in \autoref{sec:evaluatin hyperbolic space}. Negative sampling is not used. We report the HR@\{1,5,10,15,20\} and NDCG@\{1,5,10,15,20\}. We set $\lambda$ to be 0.1, and $m_{item}$ and $m_{so}$ are tuned from 0 to 40 respectively.

Second, we compare HSCML with other baseline models. To keep consistent with the settings in \cite{wang2020relational}, we set the latent size to be 128 and report HR@\{10,20\} \& NDCG@\{10,20\}. Because they used negative sampling in their experiments, we also use negative sampling here. We randomly sample 999 negative samples for each user and rank them together with the test item. For HSCML, we set the $\lambda$ to be 0.1. We tune $m_{item}$ and $m_{so}$ from 0 to 40 respectively. We repeat the experiment 10 times and report the average. Other parameters and settings are the same as we did in \autoref{sec:evaluatin hyperbolic space} when evaluating the performance of hyperbolic space. The numbers of all methods except for SAMN and HSCML come from the paper of RML-DGATs \cite{wang2020relational}. We do so because the Ciao and Epinions they used are slightly different from ours, and we are unable to recover their datasets. Moreover, the implementation of RML-DGATs is not provided, so it is difficult for us to try our datasets on their model. What we can do is to run our datasets on the implementation of SAMN\footnote{\url{https://github.com/chenchongthu/SAMN}} and report the numbers we get. Hopefully this can provide an insight on how hyperbolic space performs compared with state-of-the-art Euclidean methods.
\subsection{Experimental Results}
\subsubsection{\textbf{SCML vs. HSCML}}
\autoref{fig:SCML} shows the results of SCML and it hyperbolic version HSCML. The conclusions in \autoref{sec:evaluatin hyperbolic space} still hold here. For example, because SCML is a distance model, and the densities of Epinions and Ciao are low as shown in \autoref{table:datasets statistics}, so it is reasonable that the performance of hyperbolic space clearly outperforms Euclidean space on both two datasets. Moreover, as the latent size increases, the performance of Euclidean space gradually catches up with hyperbolic space, which is consistent with our conclusion regarding the influence of the latent size.

\subsubsection{\textbf{HSCML vs. baseline approaches}}
We compare the performance of HSCML with other baseline methods. The results are shown in \autoref{table:performance comparison}. HSCML has a comparable performance with the state-of-the-art embedding methods such as SAMN and RML-DGATs, while outperforms other baseline approaches. This suggests that, hyperbolic embeddings can easily reach a comparable or better performance than existing baseline approaches with a simple distance model, whereas the state-of-the-art baselines such as SAMN and RML-DGATs are much more complicated because of the memory modules and GAT units. This demonstrates the superiority of hyperbolic space over Euclidean space. Future works to design a customized model for hyperbolic space may produce a model that outperforms all existing approaches.

%% file: related-work.tex
\section{related work}
\label{sec_related_work}
Hyperbolic space has always been a popular research domain in Mathematics \cite{cannon1997hyperbolic}. Some works have been done to explore the tree-like structure of graphs \cite{adcock2013tree, chen2013hyperbolicity} and the relations between hyperbolic space and hierarchical data such as languages and complex networks \cite{krioukov2010hyperbolic, nickel2017poincare}. Such works have demonstrated the consistency between real-world scale-free and hierarchical data and the hyperbolic space, providing theoretical basis for recent works which apply hyperbolic space to various tasks including link prediction, node classification, and recommendation.

Some researchers apply hyperbolic space to traditional metric learning approaches such as HyperBPR \cite{vinh2018hyperbolic} and HyperML \cite{vinh2020hyperml}. Some try to adopt hyperbolic space to neural networks and define hyperbolic neural network operations, producing powerful models such as hyperbolic neural networks \cite{ganea2018hyperbolic}, hyperbolic graph neural networks \cite{liu2019hyperbolic} and hyperbolic convolutional neural networks \cite{chami2019hyperbolic}. Meanwhile, \cite{chamberlain2019scalable} provides a scalable hyperbolic recommender system for industry use. \cite{wang2019hyperbolic} applies hyperbolic space to heterogeneous networks for link prediction task. \cite{feng2020hme} applies hyperbolic space to next-POI recommendation. \cite{papadis2017path} proposes a path-based recommendation approach with hyperbolic embeddings, etc.